\documentclass[journal]{IEEEtranTCOM}
\normalsize

\usepackage[utf8]{inputenc}
\usepackage{amsmath}
\usepackage{tikz}
\usepackage{amssymb}
\usepackage{latexsym}
\usepackage{ifpdf}
\usepackage{wasysym}
\usepackage{algorithm, algorithmic}
\usepackage{balance}
\usepackage{cite}

\newtheorem{lemma}{Lemma}

\newtheorem{corollary}{Corollary}
\newcommand{\Ei}{\mbox{Ei}}

\begin{document}
\title{Optimizing User Selection Schemes in Vector Broadcast Channels}
\author{Meng~Wang~\IEEEmembership{Student Member,~IEEE},
		Tharaka Samarasinghe~\IEEEmembership{Member,~IEEE} and 
        Jamie~S.~Evans~\IEEEmembership{Member,~IEEE}
\thanks{M. Wang is with the Dept. of Electrical and Electronic Engineering, University of Melbourne, Australia (email: meng.wang@unimelb.edu.au).}%
\thanks{T. Samarasinghe and J. S. Evans are with the Dept. of Electrical and Computer Systems Engineering, Monash University, Australia (email: \{tharaka.samarasinghe, jamie.evans\}@monash.edu).}%
}


\maketitle

\vspace{-2cm}
\begin{abstract}
In this paper, we focus on the ergodic downlink sum-rate performance of a system consisting of a set of heterogeneous users. We study three user selection schemes to group near-orthogonal users for simultaneous transmission. The first scheme is a random selection policy that achieves fairness, but does not exploit multi-user diversity. The second scheme is a greedy selection policy that fully exploits multi-user diversity, but does not achieve fairness, and the third scheme achieves fairness while partially exploiting multi-user diversity. 
We also consider two beamforming methods for data transmission, namely, maximum-ratio transmission and zero-forcing beamforming. In all scheduling schemes studied in the paper, there is a key parameter that controls the degrees of orthogonality of channel directions between co-scheduled users. We focus on optimally setting this parameter for each scheduling scheme such that the ergodic downlink sum-rate is maximized. To this end, we derive analytical expressions for the ergodic downlink sum-rate considering each scheduling scheme. Numerical results are also presented to provide further insights. 
\end{abstract}

\section{Introduction}
\PARstart{I}{n} the past decade, multi-user Multiple-Input Multiple-Output (MU-MIMO) systems have sparked a great deal of interest, and have become one of the foundation technologies for the next generation wireless communication networks such as LTE-Advanced~\cite{lte}. In these systems, channel-dependent user selection refers to the process of selecting a subset of users according to their channel conditions to serve simultaneously on a time/frequency resource. The user selection problem is combinatorial in nature and obtaining the optimal scheduled user set requires an exhaustive search over all possible combinations of the users. However, the computational complexity of this exhaustive search is too high for most practical implementations. 

It is well known that transmitting to a set of nearly orthogonal users is close to optimal in terms of sum-rate~\cite{Whiting2007}. Therefore, many practical scheduling algorithms have been proposed in recent years that tradeoff performance for the reduction of complexity associated with user selection. The authors in~\cite{vtc2013} have categorized them into two frameworks, namely, the greedy scheduling framework and the opportunistic scheduling framework. In the greedy framework, the base station (BS) first selects a subset of users sequentially such that their channel directions are close to orthogonal. Beamforming vectors are then constructed based on the channel directions of the selected users. In the opportunistic framework, the BS first randomly constructs a number of mutually orthogonal beams, and then selects the users with the highest signal-to-interference-plus-noise-ratio (SINR) on each beam for communication~\cite{Hassibi2005, TharakaIT}. 

The earliest work on greedy user scheduling appeared in~\cite{Goldsmith2005_Globecom}, where near-orthogonal users having the strongest channel magnitudes were selected to guarantee a high SINR and gain through multi-user diversity. The authors in~\cite{Goldsmith2006} have shown that the semi-orthogonal user selection (SUS) scheme with zero-forcing beamforming (ZF-BF) achieves a sum-rate close to dirty paper coding (DPC) asymptotically as the number of users grows large. In these greedy scheduling schemes, there is an important parameter $\alpha\in[0,1]$ that controls the degree of orthogonality between co-scheduled users' channel directions, \emph{i.e.}, two users are considered to be near-orthogonal (hence can be co-scheduled together) if the inner product between their normalized channel vectors is less than $\alpha$. Choosing a suitable value for $\alpha$, which is the main focus of this paper, is known to be a difficult problem~\cite{Min2009, AusCTW2013, Sun2010, Song2012, Trivellato2007}, as the optimal $\alpha$ that maximizes the ergodic sum-rate depends on the noise power, the number of users in the system, and the beamforming method used by the BS. If $\alpha$ is too small, the BS may not be able to find any co-scheduled users. On the other hand, if $\alpha$ is too large,  severe inter-user interference may occur between the scheduled users due to poor orthogonality. Finding the optimal $\alpha$ that maximizes the ergodic downlink sum-rate is not straightforward. The uncertainty and difficulty of choosing an optimal $\alpha$ has in fact led to many alternative greedy user scheduling methods that avoid the necessity of $\alpha$ (\emph{e.g.}, \cite{Song2012, Trivellato2007, Dimic2005, Leung2013, pimrc2012, Ozyurt2012}). 

In order to find the optimal $\alpha$, an analytical expression of the exact ergodic sum-rate is required. Obtaining the exact ergodic sum-rate expression for the most general scenario is difficult. Therefore, previous works focusing on analyzing the ergodic sum-rate of semi-orthogonal users have either assumed that the number of users is large~\cite{Whiting2007, Goldsmith2006, Sun2010, Dimic2005, Zoltowski2008, Min2013TC}, or the BS has only two transmit antennas (or alternatively, only schedule up to a pair of users)~\cite{Ozyurt2012, Zhang2007, Lu2009}. 
In particular, the authors in~\cite{Lu2009} have considered a two transmit antenna system, and analyzed the ergodic sum-rate of the SUS scheme with ZF-BF. 

In all of the aforementioned works, the scheduling algorithms are designed and analyzed with the assumption that users are homogeneous, \emph{i.e.}, the large-scale fading gain value (may alternatively referred to as the path loss in the paper) of each user is equal to one. Our work is mainly different from these works as we consider a more practical setup where the path loss (PL) values between each user and the BS are different. Recently, analysis on the ergodic sum-rate of semi-orthogonal users considering the heterogeneous user setting can be found in~\cite{Sohn2010, Min2013CL}. In~\cite{Sohn2010}, the authors have derived the sum-rate scaling law of the SUS scheme with ZF-BF, as the number of users grows large. In~\cite{Min2013CL}, the authors have extended the result to a more general PL model. In this heterogeneous user setting, we focus on finding the optimal $\alpha$ that maximizes the ergodic downlink sum-rate.

Similar to~\cite{Lu2009}, we also assume that the BS has two transmit antennas for analytical tractability\footnote{The difficulty of obtaining the exact ergodic sum-rate expression arises when selecting more than two users. A two transmit antenna system is sufficient for the analysis of scheduling up to a pair of users.} (the two antenna configuration is also supported in practice, for example, transmission modes 2-9 in LTE~\cite{lte}). We consider three different schemes to schedule the heterogeneous users. In the first scheme, the users are selected randomly regardless of their channel magnitudes, as long as their channel directions are near-orthogonal. In this scheme, fairness is naturally achieved, however, since the users are selected without considering their channel magnitudes, multi-user diversity is not exploited. In the second scheme, near-orthogonal users with the highest received power are selected, \emph{i.e.}, the scheduler takes the PL values, channel magnitudes and channel direction information of the users into account. In this scheme, fairness is not achieved, however, multi-user diversity is fully exploited. The third scheme is based on a CDF-based scheduling policy in~\cite{Park2005}. In this scheme, near-orthogonal users with the highest received power relative to its own statistics are selected. As a result, fairness is achieved and multi-user diversity is partially exploited. We name the three schemes as the Random User Selection (RUS) scheme, the Max-Gain User Selection (MUS) scheme, and the CDF-Based User Selection (CUS) scheme, respectively. For each scheduling scheme, we also consider two beamforming methods for data transmission. To this end, we consider the ZF-BF as in~\cite{Lu2009}. In addition, we also consider maximum ratio transmission (MRT), which is more difficult to analyze due to the presence of inter-user interference. 

Having introduced the above scheduling schemes and beamforming methods, we focus on optimally setting $\alpha$ for each scenario such that the ergodic downlink sum-rate is maximized. To this end, we derive analytical expressions for the exact ergodic sum-rate considering each scheduling scheme and beamforming method. We also show how the ergodic sum-rate expressions for the homogeneous user setting can be easily deduced using our results. Numerical results are presented to provide further insights on the performance of the above scheduling schemes and the behaviour of $\alpha^\star$, which is the optimal $\alpha$ that maximizes the ergodic downlink sum-rate. Our results show that for a range of practical parameters, $\alpha^\star$ decreases with the number of users in the system. Moreover, when using MRT, $\alpha^\star$ decreases with transmit signal-to-noise-ratio (SNR) for all schemes; and when using ZF-BF, $\alpha^\star$ increases with transmit SNR. 

The organization of the paper is as follows. The system model is presented in Section~\ref{sec:ch3_sys_model}. The detailed operation of each scheduling scheme is outlined in Section~\ref{sec:ch3_algorithms}. Analysis of the RUS scheme, MUS scheme and the CUS scheme is presented in Sections~\ref{sec:RUS}, \ref{sec:SUS} and \ref{sec:CUS}, respectively. In Section~\ref{sec:ch3_results}, we present numerical results, and Section~\ref{sec:ch3_conclusion} concludes the paper.

%
%
%
\section{System Model}\label{sec:ch3_sys_model}
We consider a single cell multi-user MISO vector broadcast channel. The base station (BS) is equipped with two transmit antennas. There are $K$ users in the network and each user is equipped with a single receive antenna. Let $\mathbf{h}_k$ denote the $2\times 1$ channel gain vector between user $k$ and the BS. The elements in $\mathbf{h}_k$ are independent and identically distributed (i.i.d.) random variables, each of which is drawn from a zero mean and unit variance circularly symmetric complex Gaussian distribution $\mathcal{CN}(0,1)$. The large-scale fading gain (alternatively referred to as path loss in the paper) between user $k$ and the BS is denoted by $g_k$. Moreover, we assume a quasi-static block fading model over time~\cite{Tse_book}. 

The BS picks a maximum of two semi-orthogonal users for data transmission. When the BS transmits to two users, we call this the multi-user transmission mode (MU-Mode). The basic criteria of selecting the two users is that they satisfy the semi-orthogonality (SO) constraint, 
\begin{align*} 
	\frac{|{\mathbf{h}}_{\pi(1)}^{\textrm{H}} \mathbf{h}_{\pi(2)}|^2}{\|\mathbf{h}_{\pi(1)}\|^2 \|\mathbf{h}_{\pi(2)}\|^2} \leq \alpha^2,
\end{align*}
where $\pi(1)$ and $\pi(2)$ denote the indices of the first and second selected users, respectively. When the BS cannot find a semi-orthogonal user to $\pi(1)$, the BS will only transmit to one user, and we call this the single-user transmission mode (SU-Mode). For the system in consideration, the probability of finding a user semi-orthogonal to $\pi(1)$ reduces to $\alpha^2$~\cite{Goldsmith2005_Globecom}. Therefore, the probability of the BS operating under SU-Mode and MU-Mode is given by $(1-\alpha^2)^{K-1}$ and $1-(1-\alpha^2)^{K-1}$, respectively. In addition to the SO constraint, $\pi(1)$ and $\pi(2)$ will also depend on the user scheduling algorithms, which we will discuss in detail in the next section.

Let $\mathcal{U}$ denote the set of scheduled users on the same time/frequency resource element. Let $\mathbf{w}_{i}$ and $s_{i}$ denote the unit-norm beamforming vector associated with user $i\in\mathcal{U}$, and the transmitted symbol on beam $\mathbf{w}_{i}$, respectively. The BS employs either MRT or ZF beamforming technique for data transmission. For MRT, the unit-norm beamforming vector $\mathbf{w}_i$ is chosen to be in the direction of the selected user's channel, \emph{i.e.}, $\mathbf{w}_i = \tilde{\mathbf{h}}_i = \frac{\mathbf{h}_i}{\|\mathbf{h}_i\|}, ~\forall i\in\mathcal{U}$. For ZF-BF, the unit-norm beamforming vector $\mathbf{w}_i$ is chosen to be the projection of $\mathbf{h}_i$ into the orthogonal subspace of the interference, hence the zero-interference condition $\mathbf{h}_j^{\textrm{H}} \mathbf{w}_i = 0, ~\forall j \neq i,~ i,j \in \mathcal{U}$, is satisfied. The received signal at user $i\in\mathcal{U}$ can be written as 
\begin{align*}
	y_{i} = \sqrt{g_i} \mathbf{h}_{i}^{\textrm{H}} \sum_{j \in \mathcal{U}} \mathbf{w}_{j} s_{j} + n_{i},
\end{align*}
where $n_{i} \sim \mathcal{CN}(0,\sigma^2)$ is the complex Gaussian noise. We assume that $\mathbb{E}[|s_{i}|^2] = \rho$ for all $i\in\mathcal{U}$, where $\rho = P_t/|\mathcal{U}|$ is a scaling parameter to satisfy the total transmit power constraint $P_t$ at the BS. For convenience, we assume that $P_t = 2$, \emph{i.e.}, the transmit power per-antenna is one. Therefore, we have $\rho = 2$ for SU-Mode and $\rho = 1$  for MU-Mode. 

The SINR at user $i \in\mathcal{U}$ when the BS transmits using beamforming technique $l \in \{ \mbox{MRT},~\mbox{ZF}\}$ is given by
\begin{align} \label{eq:sinr_expression_general}
	\mbox{SINR}_i^l = \frac{g_i |\mathbf{h}_i^{\textrm{H}} \mathbf{w}_i|^2}{\tilde{\sigma}^2 + g_i \sum_{\substack{j\in \mathcal{U} \\ j\neq i}} |\mathbf{h}_i^{\textrm{H}} \mathbf{w}_j|^2},
\end{align}
where $\tilde{\sigma}^2 = \sigma^2/\rho$ is the effective noise power. Let $S_i = \frac{g_i \|\mathbf{h}_i\|^2}{\tilde{\sigma}^2}$. The CDF of $S_i$ is given by $F_{S_i}(x)=\gamma\left(2, \frac{\tilde{\sigma}^2}{g_i}x\right)$, where $\gamma(2,x) = 1-(x+1)\exp(-x)$ is the lower incomplete Gamma function~\cite{Abramowitz_Stegun}. 

Consider the case when $|\mathcal{U}|=2$. Let $Y^{\mbox{\tiny{MRT}}} = |\tilde{\mathbf{h}}_i^{\textrm{H}} \tilde{\mathbf{h}}_j|^2$ and $Y^{\mbox{\tiny{ZF}}} = |\tilde{\mathbf{h}}_i^{\textrm{H}} \mathbf{w}_i|^2$. For any two users, the squared normalized inner product is uniformly distributed over $[0,1]$ for the two-transmit antenna system~\cite{Whiting2007}. Therefore, $Y^{\mbox{\tiny{MRT}}}$ is uniformly distributed over $[0,\alpha^2]$ due to the SO constraint. Moreover, $Y^{\mbox{\tiny{ZF}}}$ is uniformly distributed over [$1-\alpha^2$, $1$] due to the SO constraint and the ZF-BF condition~\cite{Lu2009}. Hence, the SINR expression in (\ref{eq:sinr_expression_general}) can be re-written as $\mbox{SINR}_i^{\mbox{\tiny{MRT}}} = \left[S_i^{-1} + Y^{\mbox{\tiny{MRT}}} \right]^{-1}$ for MRT and $\mbox{SINR}_i^{\mbox{\tiny{ZF}}} = S_i Y^{\mbox{\tiny{ZF}}}$ for ZF-BF. 

Finally, the ergodic downlink sum-rate of the system  when the BS transmits using beamforming technique $l \in \{ \mbox{MRT},~\mbox{ZF}\}$ is given by
\begin{align*}
	\mathcal{R}^l_{\mbox{\small{sum}}}(\alpha) = \sum_{i\in\mathcal{U}} \mathbb{E}\left[\log(1+\mbox{SINR}^l_i)\right].
\end{align*}
The sum-rate is clearly a function of $\alpha$, and if $\alpha$ is too large, the scheduled users will have poor orthogonality, hence degrading the system performance. On the other hand, if $\alpha$ is too small, there may not be a pair that satisfy the SO constraint, hence the spatial multiplexing capability of having multiple transmit antennas is not exploited. In this paper, we focus on finding the optimal $\alpha$ that maximizes the ergodic downlink sum-rate for each scheduling scheme and beamforming technique in consideration, \emph{i.e.}, we study the following optimization problem,
\begin{align*}
	\underset{\alpha \in [0,1]}{\mbox{maximize}} & ~~ \mathcal{R}^l_{\mbox{\small{sum}}}(\alpha).
\end{align*}
To this end, we will need to first derive analytical expressions for the ergodic downlink sum-rate considering each case of interest. For convenience, we define the following terms,
\begin{align*}
	&\lambda = (1-\alpha^2)^{K-1} = \mbox{probability of operating in the SU-Mode} \\
	&\mathcal{R}_S= \mbox{ergodic rate of $\pi(1)$ under SU-Mode} \\
	&\mathcal{R}_{M1}^l(\alpha) = \mbox{ergodic rate of $\pi(1)$ under MU-Mode} \\ & \hspace{1.8cm} \mbox{and BF method}~l\in\{ \mbox{MRT},~\mbox{ZF}\} \\
	&\mathcal{R}_{M2}^l(\alpha) = \mbox{ergodic rate of $\pi(2)$ under MU-Mode} \\ & \hspace{1.8cm} \mbox{and BF method}~l\in\{ \mbox{MRT},~\mbox{ZF}\}.
\end{align*}
Thus, the objective function of the optimization problem of interest can be re-written as 
\begin{align} \label{eq:ergodic_sum_rate_3terms}
	\mathcal{R}^l_{\mbox{\small{sum}}}(\alpha) = \lambda \mathcal{R}_S + (1-\lambda) [\mathcal{R}^l_{M1}(\alpha) + \mathcal{R}^l_{M2}(\alpha)]. 
\end{align}
Before obtaining expressions for $\mathcal{R}^l_{\mbox{\tiny{sum}}}(\alpha)$, we will discuss three user selection algorithms in detail in the next section.

\section{User Selection Algorithms}\label{sec:ch3_algorithms}
In this paper, we study three different user selection schemes to schedule near-orthogonal users in a heterogeneous environment. The schemes in consideration are the Random User Selection (RUS) scheme, the Max-Gain User Selection (MUS) scheme and the CDF-based User Selection (CUS) scheme. The following subsections describe the operation of each scheme in detail. 


\subsection{Random User Selection}
In the RUS scheme, only the channel direction information is used for scheduling purposes. The idea is to co-schedule users whose channel directions are near-orthogonal, regardless of the channel magnitudes. The BS starts by randomly picking up a user as $\pi(1)$. The BS then randomly picks a user $j\neq \pi(1)$ that satisfies the SO constraint as $\pi(2)$. In this scheme, fairness is naturally achieved. However, since the users are selected without considering their channel magnitudes, multi-user diversity is not exploited. Algorithm~\ref{alg:RUS} describes the RUS scheme. 

\begin{algorithm}[ht!]
\begin{algorithmic}
	\STATE STEP 1: Re-order the users randomly, let $\{(1),\ldots,(K)\}$ denote the randomized indices
	\STATE STEP 2: $\pi(1) = (1)$
	\FOR {$j=2,\ldots,K$}
	\IF {$\left(\frac{|{\mathbf{h}}_{\pi(1)}^{\textrm{H}} {\mathbf{h}}_{(j)}|^2}{\|\mathbf{h}_{\pi(1)}\|^2 \|\mathbf{h}_{(j)}\|^2} \leq \alpha^2\right)$}
		\STATE STEP 3: $\pi(2) = j$
		\STATE break
	\ENDIF
	\ENDFOR
\end{algorithmic}
\caption{Random User Selection (RUS) Scheme}
\label{alg:RUS}
\end{algorithm}

\subsection{Max-Gain User Selection}
Compared to the RUS scheme, the MUS scheme takes the PL value, channel magnitude and channel direction information of the users into account. In the MUS scheme, the BS picks $\pi(1)$ such that $\pi(1) = \arg\max_{1\leq k\leq K} g_k \|\mathbf{h}_k\|^2$. Then, the BS picks $\pi(2)$ such that $\pi(2) = \arg\max_{j\in\mathcal{X}} g_j \|\mathbf{h}_j\|^2$, where $\mathcal{X}$ denotes the set of users semi-orthogonal to $\pi(1)$. That is, the user with the highest received power is chosen as $\pi(1)$, and the user  with the highest received power among all users satisfying the SO constraint is chosen as $\pi(2)$. As a consequence, fairness will not be achieved since the user with a small PL value will be less likely to be selected compared to a user with a large PL value. However, multi-user diversity will be fully exploited.  Algorithm~\ref{alg:SUS} describes the MUS scheme. 
\begin{algorithm}	
\begin{algorithmic}[ht!]
	\STATE STEP 1: Rank the users according to \\~~~~~~~~~
	$g_{(1)} \|\mathbf{h}_{(1)}\|^2 >g_{(2)} \|\mathbf{h}_{(2)}\|^2 > \cdots > g_{(K)} \|\mathbf{h}_{(K)}\|^2$
	\STATE STEP 2: $\pi(1) = (1)$ 
	\FOR {$j = 2\ldots, K$}
		\IF {$\left(\frac{|{\mathbf{h}}_{\pi(1)}^{\textrm{H}} {\mathbf{h}}_{(j)}|^2}{\|\mathbf{h}_{\pi(1)}\|^2 \|\mathbf{h}_{(j)}\|^2} \leq \alpha^2\right)$}
			\STATE STEP 3: $\pi(2) = (j)$
			\STATE break
		\ENDIF
	\ENDFOR
\end{algorithmic}
\caption{Max-Gain User Selection (MUS) Scheme}
\label{alg:SUS}
\end{algorithm}

\subsection{CDF-Based User Selection}
Finally, the CUS scheme exploits multi-user diversity gain (to a certain extent) while guaranteeing scheduling fairness among the users. In the CUS scheme, the BS will utilize the distribution of $S_k$. The BS first performs the following transformation
\begin{align*}
	p_k = F_{S_k}\left( \frac{g_k}{\tilde{\sigma}^2} \|\mathbf{h}_k\|^2 \right) = \gamma\left(2, \|\mathbf{h}_k\|^2 \right),~~ k=1,\ldots,K.
\end{align*}
We note that the effect of the PL values are removed when the above transformation is performed. This means, the $g$'s are set to one when taking the scheduling decision. However, although the $g$'s can be set to one for the scheduling decision, they have to be considered when the ergodic sum-rate is calculated. Therefore, the PL values cannot be simply set to one when analyzing the performance measure of interest. In practice, the BS only needs the knowledge of the channel magnitude of each user for this transformation.

The BS picks $\pi(1)$ such that $\pi(1) = \arg\max_{1\leq k\leq K} p_k$. Then, the BS picks $\pi(2)$ such that $\pi(2) = \arg\max_{j\in\mathcal{X}} p_j$, where $\mathcal{X}$ denotes the set of users semi-orthogonal to $\pi(1)$. That is, the user with the highest received power relative to its own statistics is chosen as $\pi(1)$, and the user with the highest received power relative to its own statistics among all users satisfying the SO constraint is chosen as $\pi(2)$. The above transformation eliminates the effect of the PL value on the scheduling decision, and each user is equally likely to be selected. As a result, fairness is guaranteed among the users. Algorithm~\ref{alg:CUS} describes the CUS scheme. 

\begin{algorithm}[ht!]
\caption{CDF-Based User Selection (CUS) Scheme}
\label{alg:CUS}
\begin{algorithmic}
	\FOR {$k=1,\ldots,K$}
		\STATE STEP 1: COMPUTE $ p_k = F_{S_k}\left( \frac{\rho g_k}{\sigma^2} \|\mathbf{h}_k\|^2 \right)$
	\ENDFOR
	\STATE STEP 2: Rank the users according to \\ ~~~~~~~~~~~$p_{(1)} > p_{(2)} > \cdots > p_{(K)}$
	\STATE STEP 3: $\pi(1) = (1)$
	\FOR {$j=2,\ldots,K$}
		\IF {$\left(\frac{|{\mathbf{h}}_{\pi(1)}^{\textrm{H}} {\mathbf{h}}_{(j)}|^2}{\|\mathbf{h}_{\pi(1)}\|^2 \|\mathbf{h}_{(j)}\|^2} \leq \alpha^2\right)$}
			\STATE STEP 4: $\pi(2) = (j)$
			\STATE break
		\ENDIF
	\ENDFOR
\end{algorithmic}
\end{algorithm}
\setlength{\textfloatsep}{0.3cm}

%
%
%

\section{Analysis for the RUS Scheme} \label{sec:RUS}
In this section, we derive expressions for the ergodic downlink sum-rate of the RUS scheme. Firstly, we consider the case where the BS operates under SU-Mode, and derive an analytical expression for the ergodic rate (which is the same for MRT and ZF-BF because there is no interference). Then, we will consider the case where the BS operates under MU-Mode, and derive expressions for the ergodic rate of each selected user by considering MRT and ZF-BF, respectively. To this end, the ergodic rate under SU-Mode is formally presented in the following lemma. For convenience, we let 
\begin{align*}
	G(x) = \exp(x) \Ei(-x)
\end{align*}
throughout this section, where $\Ei(x) = - \int_x^\infty \frac{e^{-t}}{t} dt$ is the exponential integral function. 
\begin{lemma}\label{lem:RUS_SUMode}
Consider a BS operating in SU-Mode by using the RUS scheme for user selection. Then, the ergodic rate of the selected user is given by
\begin{align*} 
\mathcal{R}_S =1 + \frac{1}{K} \sum_{k=1}^K G\left(\frac{\tilde{\sigma}^2}{g_{k}}\right) \left(\frac{\tilde{\sigma}^2}{g_k}-1\right), 
\end{align*}
for both MRT and ZF-BF.
\end{lemma}
\begin{IEEEproof}
Let $X=\log(1+S_k)$, we have $F_X(x) = F_{S_k}(e^x-1)$. The ergodic rate of user $k$ can be written as
\begin{align}
	\mathbb{E} [X]	& = \int_0^\infty [1-F_X(x)] dx = \int_0^\infty \left[1 - F_{S_{k}}(e^x-1)\right] dx   \nonumber \\
								&=  \int_0^\infty  \frac{\left[\frac{\tilde{\sigma}^2 u}{g_k}+1\right] \exp\left(-\frac{\tilde{\sigma}^2 u}{g_k}\right)}{u+1} du \nonumber \\
								&= 1 + G\left(\frac{\tilde{\sigma}^2}{g_{k}}\right) \left(\frac{\tilde{\sigma}^2}{g_{k}}-1\right), \label{eq:RUS_SUMode_conditional}
\end{align}
where $u = \exp(x)-1$. 
Since each user is equally likely to be selected in the RUS scheme, averaging (\ref{eq:RUS_SUMode_conditional}) over $k$ completes the proof. 
\end{IEEEproof}

Next, we will focus on the case where the BS operates in MU-Mode. Since we have assumed that the transmit power per antenna is one, we have $\rho= 1$ for the MU-Mode. Unlike the SU-Mode, the ergodic rate of the selected users will be different for MRT and ZF-BF, and we will study the two cases separately. The ergodic rate of each selected user when the BS operates using MRT is formally presented in the following lemma. 
\begin{lemma} \label{lem:RUS_MUMode_MRT}
Consider a BS operating in MU-Mode by using the RUS scheme for user selection and MRT for data transmission. The ergodic rate of each selected user is given by
\begin{align*}
&\mathcal{R}_{M1}^{\mbox{\tiny{MRT}}}(\alpha) = \mathcal{R}_{M2}^{\mbox{\tiny{MRT}}}(\alpha) = \frac{1}{K} \times
\\ & \sum_{k=1}^K \left[G\left(\frac{\sigma^2}{g_k\alpha^2}\right) + \frac{1}{\alpha^2} G\left(\frac{\sigma^2}{g_k}\right) - \frac{\alpha^2+1}{\alpha^2} G\left(\frac{\sigma^2}{g_k(1+\alpha^2)}\right)\right].
\end{align*}
\end{lemma}
\begin{IEEEproof}
	See Appendix A. 
\end{IEEEproof}
Similarly, the ergodic rate of each selected user when the BS uses ZF-BF is formally presented in the following lemma. 
\begin{lemma} \label{lem:RUS_MUMode_ZF}
Consider a BS operating in MU-Mode by using the RUS scheme for user selection and ZF-BF for data transmission. The ergodic rate of each selected user is given by
\begin{align*}
\mathcal{R}_{M1}^{\mbox{\tiny{ZF}}}(\alpha) &= \mathcal{R}_{M2}^{\mbox{\tiny{ZF}}}(\alpha) = \frac{1}{K} \times \\
& \sum_{k=1}^K \left[\frac{1-\alpha^2}{\alpha^2} G\left(\frac{\sigma^2}{g_k(1-\alpha^2)}\right)  - \frac{1}{\alpha^2} G\left(\frac{\sigma^2}{g_k}\right)\right].
\end{align*}
\end{lemma}
\begin{IEEEproof}
	See Appendix A. 
\end{IEEEproof}

Finally, substituting $\mathcal{R}_S$, $\mathcal{R}^{\mbox{\tiny{MRT}}}_{M1}(\alpha)$ and $\mathcal{R}^{\mbox{\tiny{ZF}}}_{M1}(\alpha)$ given in Lemmas~\ref{lem:RUS_SUMode},~\ref{lem:RUS_MUMode_MRT} and~\ref{lem:RUS_MUMode_ZF} into equation (\ref{eq:ergodic_sum_rate_3terms}), we can obtain expressions for the ergodic downlink sum-rate of the RUS scheme.



\section{Analysis for the MUS Scheme} \label{sec:SUS}
In this section, we analyze the ergodic downlink sum-rate of the MUS scheme. Recall that in the MUS scheme, the BS picks the first user and second user such that $\pi(1) = \arg\max_{1\leq k\leq K} g_k\|\mathbf{h}_k\|^2$ and $\pi(2) = \arg\max_{j\in\mathcal{X}} g_j \|\mathbf{h}_j\|^2$. Similar to the previous section, we first consider the BS operating in the SU-Mode, and derive an analytical expression for $\mathcal{R}_S$. Then, we will consider the case where the BS operates in the MU-Mode. Firstly, the ergodic rate under the SU-Mode is formally presented in the following lemma. 
\begin{lemma} \label{lem:SUS_SUMode}
Consider a BS operating in SU-Mode by using the MUS scheme for user selection. Then, the ergodic rate of the selected user is given by
\begin{multline*}
	\mathcal{R}_S = \int_0^\infty 1 - \prod_{k=1}^K \left[1- \left((\exp(x)-1) \frac{\tilde{\sigma}^2}{g_k}+1\right) \right. \\ \left. \exp\left(- (\exp(x)-1) \frac{\tilde{\sigma}^2}{g_k}\right)  \right] dx,
\end{multline*}
for both MRT and ZF-BF. 
\end{lemma}
\begin{IEEEproof}
For the MUS scheme, the instantaneous rate under SU-Mode is given by $\log(1+ \max_{1\leq k \leq K} S_k)$. The CDF of $S_{\pi(1)} = \max_{1\leq k \leq K} S_k$ is the CDF of the maximum of $K$ independent but not identically distributed (i.n.i.d.) random variables (RVs), which is simply given by
\begin{align} \label{eq:SUS_SNR_User1}
	F_{S_{\pi(1)}}(x) = \prod_{k=1}^K \gamma\left(2, \frac{\tilde{\sigma}^2}{g_k} x\right). 
\end{align}
Therefore, the ergodic rate is given by
\begin{align*} 
\mathcal{R}_S &= \mathbb{E}[\log(1 + S_{\pi(1)})]  \\
				&= \int_0^\infty 1 -  \prod_{k=1}^K \gamma\left(2,(\exp(x)-1)\frac{\tilde{\sigma}^2}{g_k}\right) dx,
\end{align*}
which completes the proof. 
\end{IEEEproof}

Next, we will focus on the case where the BS operates in MU-Mode. Compared to the RUS scheme, the MUS scheme exploits multi-user diversity by scheduling users having higher received power. Since the simple random user selection is omitted, the ergodic rate of each selected user will be different when the BS operates in MU-Mode. Thus, we have to consider each user separately, and we start by deriving the ergodic rate of the first selected user. 
\begin{lemma} \label{lem:SUS_MRT_MUmode_1stuser}
Consider a BS operating in MU-Mode by using the MUS scheme for user selection and beamforming technique $l\in\{\mbox{MRT},~\mbox{ZF}\}$ for data transmission. The ergodic rate of the first selected user $\pi(1)$ is given by
\begin{align*} 
	\mathcal{R}_{M1}^{l}(\alpha) {=} \frac{\sigma^4}{\alpha^2} \sum_{k=1}^K \frac{1}{g_k^2} \int_0^\infty \hspace{-0.3cm} \Upsilon^{l}(x) x\exp(-\sigma^2x) \prod_{\substack{j=1 \\ j\neq k}}^K \hspace{-0.1cm} \gamma\left(2,\frac{\sigma^2}{g_j}x\right) dx,
\end{align*}
where 
\begin{multline} \label{eq:g_MRT}
	\Upsilon^{\mbox{\tiny{MRT}}}(x) {=} \log\left(\frac{1+x+\alpha^2x}{1+x}\right) + \frac{1}{x} \log\left(\frac{1+(x^{-1}+\alpha^2)^{-1}}{1+x}\right) \\ + \alpha^2 \log\left(1+(x^{-1}+\alpha^2)^{-1}\right)
\end{multline}
and
\begin{multline}\label{eq:g_ZFBF}
	\Upsilon^{\mbox{\tiny{ZF}}}(x) {=} \frac{1}{x} \Big\{(1+x) \log(1+x)  - [1+(1-\alpha^2)x] \\ \log[1+(1-\alpha^2)x] \Big\}  - \alpha^2.
\end{multline} 
\end{lemma}
\begin{IEEEproof}
	See Appendix  B. 
\end{IEEEproof}
Next, we focus on the second selected user. Obtaining an expression for the rate of the second selected user is more difficult compared to the first because $\mathcal{X}$ is a random set, and the cardinality of the set is a random number. The randomness of $\mathcal{X}$ stems on the fact that the channel vectors are random. Thus, we have to consider the maximum of $|\mathcal{X}|$ non-identically distributed random variables where $|\mathcal{X}|$ is a random variable itself. Before we present the results, we define the following matrices. First, we define $\mathbf{C}^k_{(i)}$ to be a $\binom{K-1}{K-i}$-by-$(K-i)$ matrix for which the rows of $\mathbf{C}^k_{(i)}$ consist of the combinations of $K-i$ indices chosen from the set $\{1,\ldots,k-1,k+1,\ldots,K\}$, \emph{i.e.}, the set of all indices between $1$ and $K$ except $k$. Moreover, we define $\overline{\mathbf{C}}^k_{(i)}$ to be a $\binom{K-1}{K-i}$-by-$(i-1)$ matrix for which the $j$th row is given by $\{1,\ldots,k-1,k+1,\ldots,K\}\setminus \mathbf{C}_{(i)}^k(j,:)$ where $\mathbf{C}_{(i)}^k(j,:)$ is the $j$th row of $\mathbf{C}_{(i)}^k$. For example, for $K=4$, $i=2$, and $k=1$, $\mathbf{C}^k_{(i)}$ and $\overline{\mathbf{C}}^k_{(i)}$ are given by $\mathbf{C}^1_{(2)} = [2,~3;~2,~4;~3,~4]$ and $\overline{\mathbf{C}}^1_{(2)} = [4;~3;~2]$, respectively. These ideas are formally presented in the following lemma. 
\begin{lemma}\label{lem:SUS_MRT_MUmode_2nduser}
Consider a BS operating in MU-Mode by using the MUS scheme for user selection and beamforming technique $l\in\{\mbox{MRT},~\mbox{ZF}\}$ for data transmission. The ergodic rate of the second selected user $\pi(2)$ is given by
\begin{multline*} 
\hspace{-0.3cm} \mathcal{R}^l_{M2}(\alpha) = \sum_{i=2}^K   \frac{(1-\alpha^2)^{i-2}}{1-\lambda} \sum_{k=1}^{K} \frac{\sigma^4}{g_k^2}    \int_0^\infty  \hspace{-0.2cm}    \Upsilon^{l}(x) x \exp\left(-\frac{\sigma^2}{g_k}x\right) \\ \hspace{0.4cm} \sum_{m=1}^{\binom{K-1}{K-i}} \prod_{j=1}^{K-i} \gamma\left(2, \frac{\sigma^2}{g_{\mathbf{C}_{(i)}^k(m,j)}}x\right) \prod_{j=1}^{i-1} \Gamma\left(2, \frac{\sigma^2}{g_{\overline{\mathbf{C}}_{(i)}^k(m,j)}} x\right) dx,
\end{multline*}
where $\mathbf{C}_{(i)}^k(m,j)$ and $\overline{\mathbf{C}}_{(i)}^k(m,j)$ denote the $(m,j)$th element of matrix $\mathbf{C}_{(i)}^k$ and $\bar{\mathbf{C}}_{(i)}^k$, respectively. Moreover, $\Upsilon^{\mbox{\tiny{MRT}}}(x)$ and $\Upsilon^{\mbox{\tiny{ZF}}}(x)$ are given in (\ref{eq:g_MRT}) and (\ref{eq:g_ZFBF}), respectively. 
\end{lemma}
\begin{IEEEproof}
	See Appendix B.
\end{IEEEproof}

Finally, substituting  $\mathcal{R}_S$, $\mathcal{R}_{M1}^{l}(\alpha)$, $\mathcal{R}_{M2}^{l}(\alpha)$ given in Lemmas~\ref{lem:SUS_SUMode},~\ref{lem:SUS_MRT_MUmode_1stuser}, and~\ref{lem:SUS_MRT_MUmode_2nduser} into equation (\ref{eq:ergodic_sum_rate_3terms}), we can obtain expressions for the ergodic downlink sum-rate of the MUS scheme.

\subsection{Special Case: Homogeneous Users}
In this subsection, we present a special case where the users' PL values are identical, and given by $g_1 = \cdots = g_K = g$. This is a scenario where the users are equidistant from the BS, and is the most commonly used model in the literature. In this case, much more simplified rate expressions can be obtained, and the results are presented in the following corollary. 
\begin{corollary} \label{cor:mrt_homo}
	Consider a BS using the MUS scheme for user selection and beamforming technique $l\in\{\mbox{MRT},~\mbox{ZF}\}$ for data transmission. If $g_1 = \cdots = g_K = g$, the ergodic rate of the selected user under SU-Mode is given by
\begin{align*} 
	\mathcal{R}_S &= \int_0^\infty 1 - [\gamma\left(2,(\exp(x)-1) \tilde{\sigma}^2/g \right)]^K dx.
\end{align*}
Moreover, the ergodic rate of the first and the second selected users under MU-Mode is given by
\begin{align*}
	\mathcal{R}_{M1}^l (\alpha) {=} \frac{K\sigma^4}{\alpha^2 g^2} \int_0^\infty \hspace{-0.3cm} \Upsilon^l(x) x \exp\left(- \frac{\sigma^2x}{g}\right) \left[\gamma\left(2,\frac{\sigma^2 x}{g}\right)\right]^{K-1} \hspace{-0.5cm} dx
\end{align*}
and
\begin{multline*}
\hspace{-0.2cm}	\mathcal{R}_{M2}^l (\alpha) {=} \frac{K \sigma^4}{(1-\alpha^2)(1-\lambda) g^2} \int_0^\infty \Upsilon^l(x) x \exp\left(- \frac{\sigma^2x}{g}\right) \\ \left\{ \left[1-\alpha^2 \Gamma\left(2, \frac{\sigma^2 x}{g}\right)\right]^{K-1} - \hspace{0.2cm} \left[\gamma\left(2,\frac{\sigma^2 x}{g}\right)\right]^{K-1}\right\} dx,
\end{multline*}
respectively, where $\Upsilon^{\mbox{\tiny{MRT}}}(x)$ and $\Upsilon^{\mbox{\tiny{ZF}}}(x)$ is given in (\ref{eq:g_MRT}) and (\ref{eq:g_ZFBF}), respectively. 
\end{corollary}

By using the results of Corollary~\ref{cor:mrt_homo}, the ergodic downlink sum-rate expressions for the MUS scheme can be obtained easily. Setting $g=1$ makes the ZF-BF results in Corollary~\ref{cor:mrt_homo} consistent with~\cite{Lu2009}.\footnote{This result corrects a typo in~\cite{Lu2009}, equation (13).}  



\section{Analysis for the CUS Scheme}\label{sec:CUS}
In this section, we analyze the ergodic downlink sum-rate of the CUS scheme. Similar to the previous sections, we consider the SU-Mode and the MU-Mode separately. The ergodic rate of the CUS scheme under SU-Mode is formally presented in the following lemma.
\begin{lemma} \label{lem:CUS_SUMode}
Consider a BS operating in SU-Mode by using the CUS scheme for user selection. Then, the ergodic rate of the selected user is given by
\begin{align} \label{eq:CUS_SUMode}
\mathcal{R}_S {=} \sum_{k=1}^K \frac{\tilde{\sigma}^4}{g_k^2} \int_0^\infty \hspace{-0.2cm} \log(1+x) x\exp\left(-\frac{\tilde{\sigma}^2}{g_k}x\right) \hspace{-0.1cm} \left[\gamma\left(2,\frac{\tilde{\sigma}^2}{g_k} x\right)\right]^{K-1} \hspace{-0.6cm}  dx.
\end{align}
\end{lemma}
\begin{IEEEproof}
For the CUS scheme, the instantaneous rate of the first selected user is given by $\log(1+S_{\pi(1)})$. The distribution of $S_{\pi(1)}$ can be derived as 
\begin{align*}
F_{S_{\pi(1)}} (x) &= \Pr\{ S_{\pi(1)} \leq x \} 
								= \sum_{k=1}^K \Pr\{ S_k \leq x, \pi(1) = k \} \\
								&= \sum_{k=1}^K \Pr\{ p_k \leq F_{S_k}(x), p_j\leq p_k, \forall j\neq k  \} \\
								&= \sum_{k=1}^K \int_0^{F_{S_k}(x)} \hspace{-1cm} \Pr\{p_k = t\} \Pr\{p_j \leq t, \forall j\neq k | p_k=t\} dt.
\end{align*}
Since $p_1,\ldots,p_K$ are i.i.d. and uniformly distributed over $[0,1]$, we have $\Pr\{p_k=t\} = 1$ and 
\begin{align*}
\Pr\{p_k = t\} \Pr\{p_j \leq t, \forall j\neq k | p_k=t\} &= \prod_{j\neq k} \Pr\{p_{j}\leq t \}  \\ &= t^{K-1}.
\end{align*}
Therefore, the distribution of $S_{\pi(1)}$ is given by
\begin{align} \label{eq:CDF_of_SNR_1stuser_CUS}
	F_{S_{\pi(1)}}(x) = \sum_{k=1}^{K}  \int_0^{F_{S_k}(x)} t^{K-1} dt = \frac{1}{K} \sum_{k=1}^K [F_{S_k}(x)]^K.
\end{align}
Then, by using (\ref{eq:CDF_of_SNR_1stuser_CUS}), the ergodic rate of the CUS scheme under SU-Mode can be calculated as
\begin{align*}
\mathcal{R}_S &{=} \int_0^\infty \log(1+x) dF_{S_{\pi(1)}}(x) \nonumber \\
											&{=} \sum_{k=1}^K \frac{\tilde{\sigma}^4}{g_k^2} \int_0^\infty \hspace{-0.2cm} \log(1+x) x \exp\left(-\frac{\tilde{\sigma}^2}{g_k}x\right) \left[\gamma\left(2,\frac{\tilde{\sigma}^2}{g_k}x\right)\right]^{K-1} \hspace{-0.5cm} dx,
\end{align*}
which completes the proof.
\end{IEEEproof}

Next, we will focus on the MU-Mode. Similar to the MUS scheme, the ergodic rate of each selected user is different to each other when the BS operates in MU-Mode. To this end, we start by deriving an expression for the ergodic rate of the first selected user, which is given in the following lemma. 
\begin{lemma} \label{lem:CUS_User1_MRT}
Consider a BS operating in MU-Mode by using the CUS scheme for user selection and beamforming technique $l\in\{\mbox{MRT},~\mbox{ZF}\}$ for data transmission. Then, the ergodic rate of the first selected user $\pi(1)$ is given by
\begin{align*} 
\mathcal{R}_{M1}^l(\alpha) 
		 &{=} \frac{1}{\alpha^2} \sum_{k=1}^{K} \frac{\sigma^4}{g_k^2} \int_0^\infty \hspace{-0.3cm} \Upsilon^{l}(x) \hspace{-0.1cm} \left[\gamma\left(2,\frac{\sigma^2x}{g_k}\right)\right]^{K-1} \hspace{-0.6cm} x\exp\left(-\frac{\sigma^2x}{g_k}\right)  dx,
\end{align*}
where $\Upsilon^{\mbox{\tiny{MRT}}}(x)$ and $\Upsilon^{\mbox{\tiny{ZF}}}(x)$ are given in (\ref{eq:g_MRT}) and (\ref{eq:g_ZFBF}), respectively.
\end{lemma}
\begin{IEEEproof}
	See Appendix C.
\end{IEEEproof}
The ergodic rate of the second selected user is given in the following lemma. 
\begin{lemma} \label{lem:CUS_MRT_User2}
Consider a BS operating in MU-Mode by using the CUS scheme for user selection and beamforming technique $l\in\{\mbox{MRT},~\mbox{ZF}\}$ for data transmission. Then, the ergodic rate of the second selected user $\pi(2)$ is given by
\begin{align*} 
\mathcal{R}_{M2}^{l}(\alpha) = \sum_{m=1}^{K-1} c_m \sum_{k=1}^{K} \frac{\sigma^4}{g_k^2} \int_0^\infty \Upsilon^l(x) \Psi_{m,k}(x) dx,
\end{align*}
where 
$c_m = \binom{K-2}{m-1} \frac{K-1}{m} \frac{\alpha^{2m} (1-\alpha^2)^{K-1-m}}{1- \lambda}$,
and $\Psi_{m,k}(x)$ is given by (\ref{eq:Psi_nk}) on top of the page, 
\begin{figure*}[t]
\begin{align} \label{eq:Psi_nk}
	\Psi_{m,k}(x) = \left[\gamma\left(2,\frac{\sigma^2}{g_k} x\right)\right]^{K-1} + \frac{m \left[\gamma\left(2,\frac{\sigma^2}{g_k} x\right)\right]^{m-1} - K\left[\gamma\left(2,\frac{\sigma^2}{g_k} x\right)\right]^{K-1}}{K-m}
\end{align} 
\hrule
\end{figure*}
and $\Upsilon^{\mbox{\tiny{MRT}}}(x)$ and $\Upsilon^{\mbox{\tiny{ZF}}}(x)$ is given in (\ref{eq:g_MRT}) and (\ref{eq:g_ZFBF}), respectively.
\end{lemma}
\begin{IEEEproof}
	See Appendix C.
\end{IEEEproof}

Finally, substituting  $\mathcal{R}_S$, $\mathcal{R}_{M1}^l(\alpha)$, and $\mathcal{R}_{M2}^{l}(\alpha)$ given in Lemmas~\ref{lem:CUS_SUMode},~\ref{lem:CUS_User1_MRT}, and~\ref{lem:CUS_MRT_User2} into equation (\ref{eq:ergodic_sum_rate_3terms}), we can obtain expressions for the ergodic downlink sum-rate of the CUS scheme.

\section{Numerical Results} \label{sec:ch3_results}
In this section, we present our numerical results. We define the transmit SNR as  $\frac{P_t}{\sigma^2}$. The PL values $\mathbf{g}= [g_1,\ldots,g_K]$ are generated according to $g_k = (d_k/d_0)^{-4}$, where $d_k$ denotes the distance between user $k$ and the BS and $d_0$ is a reference distance for the antenna far field~\cite{Goldsmith05}. For simplicity, we assume $d_0=1$ kilometre (km). We assume a specific scenario for the user locations where we evenly place the users between $0.5$ to $1.5$ km.\footnote{We start at $0.5$ km to avoid a very large PL value when the distance is too small.}

\begin{figure}[t]
\centering
\includegraphics[width=3.6in]{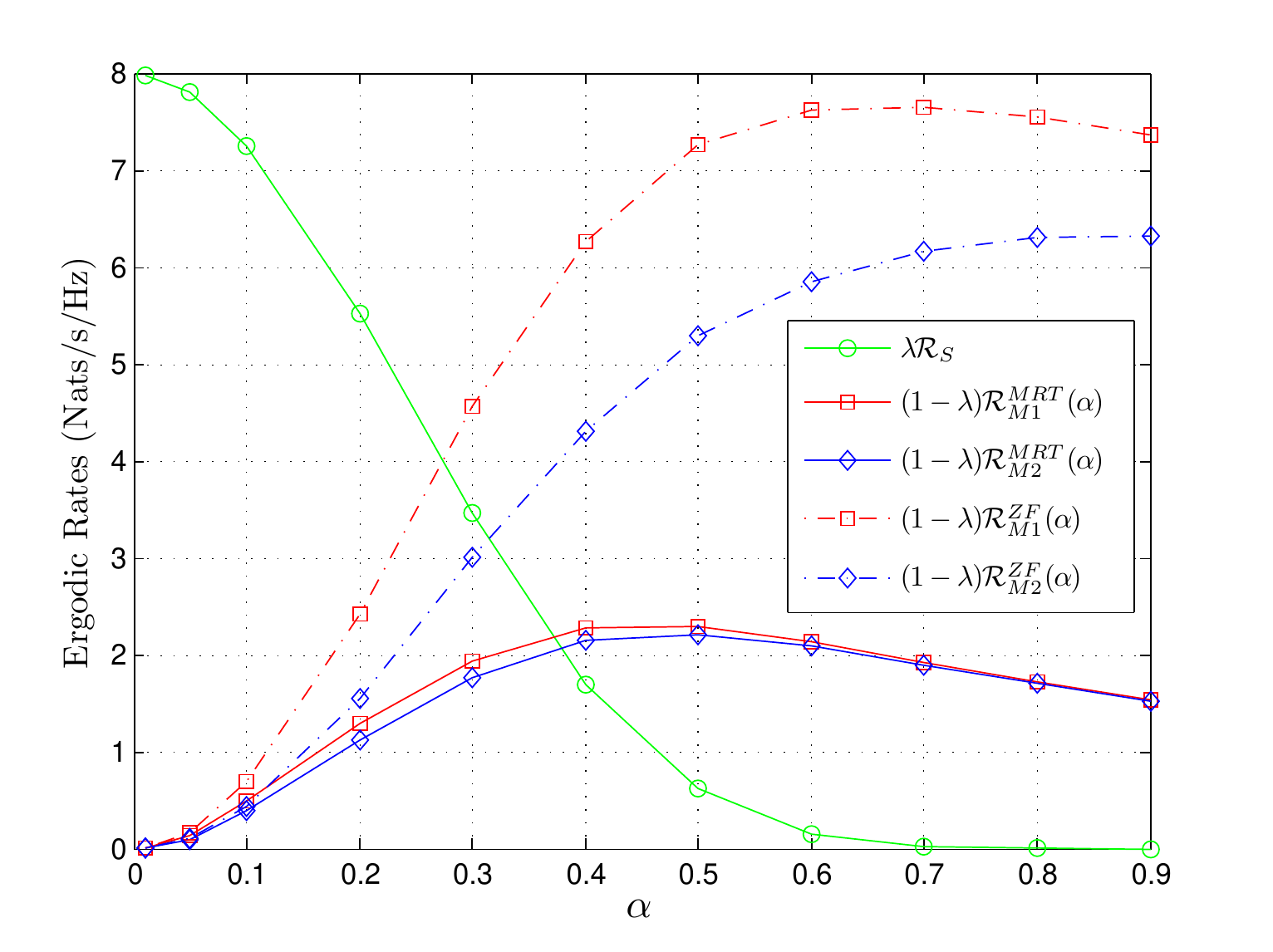}
\vspace{-0.3cm}
\caption{Ergodic Sum-Rate Components vs. $\alpha$ for the MUS Scheme, where SNR = 20dB and $K=10$.}
\label{fig:all_3_R}
\includegraphics[width=3.5in]{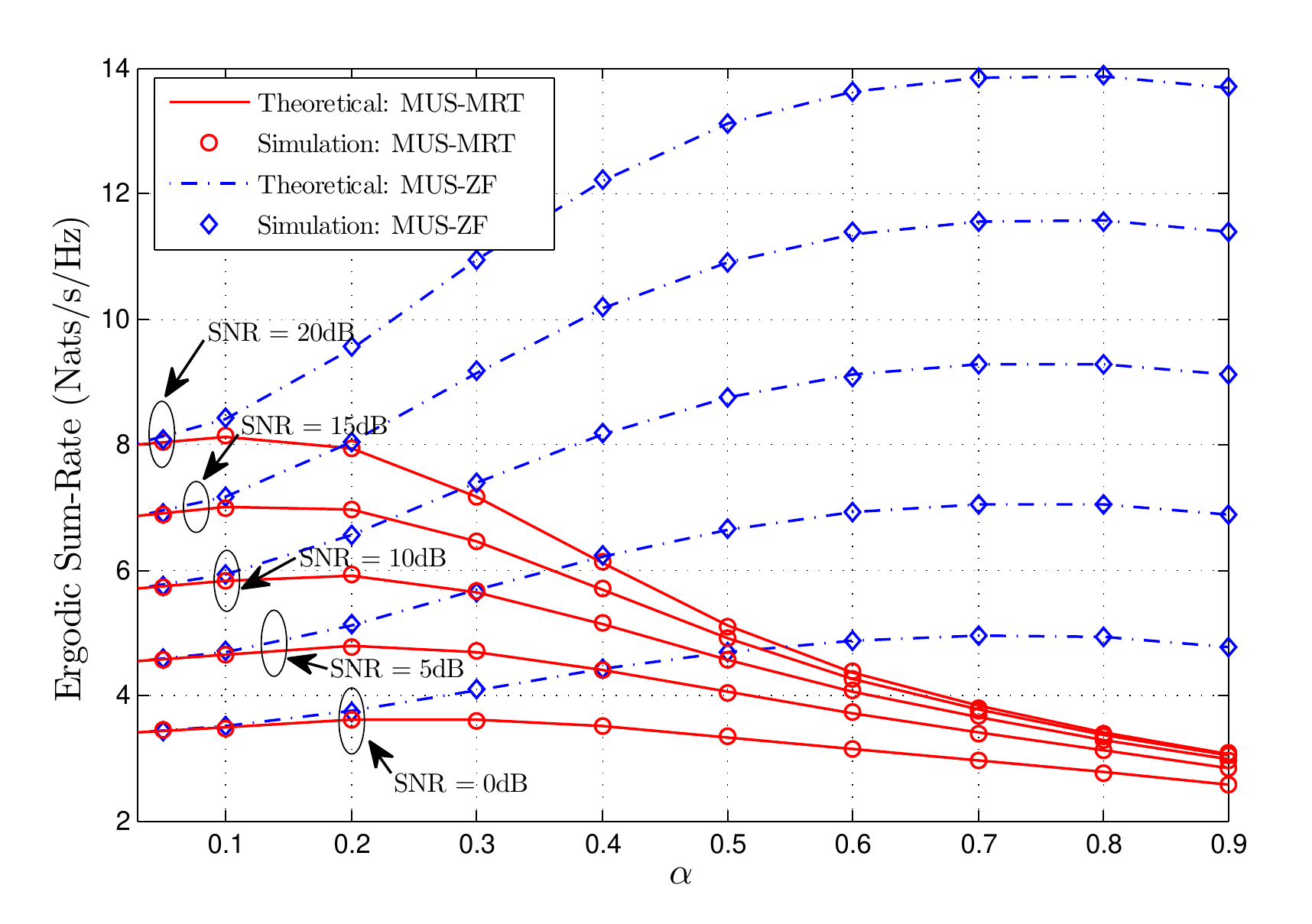}
\vspace{-1cm}
\caption{Ergodic Sum-Rate vs. $\alpha$ for the MUS Scheme, where $K=10$.}
\label{fig:SUS_K10}
\end{figure}

Figure~\ref{fig:all_3_R} shows the three components of the ergodic sum-rate in equation (\ref{eq:ergodic_sum_rate_3terms}) as $\alpha$ changes, using the MUS scheme. The total number of users is set at $K=10$, and the transmit SNR is set to 20dB. As shown in the figure, the ergodic rate in SU-Mode decreases as we increase $\alpha$, since the probability of operating in SU-Mode decreases with $\alpha$. The behaviour of the ergodic rate of each selected user in MU-Mode is more interesting. Firstly, the ergodic rate of the first selected user is always larger than the second selected user due to the MUS algorithm. The effect is greater for ZF-BF because there is no interference. Secondly, we can clearly observe the fundamental tradeoff between the following two scenarios: 1) the BS not being able to find a semi-orthogonal user to operate in MU-Mode (\emph{i.e.}, $\alpha$ too small), and 2) poor orthogonality between the selected users leading to a lower rate (\emph{i.e.}, $\alpha$ too large). As a result, the ergodic rate of each selected user in MU-Mode increases with $\alpha$ initially, and as we further increase $\alpha$, the rate will start to decrease. The ergodic downlink sum-rate will be the sum of the three rates in the figure, and is presented in Fig.~\ref{fig:SUS_K10}. 

\begin{figure}[t]
\centering
	\centering
	\includegraphics[width=3.45in]{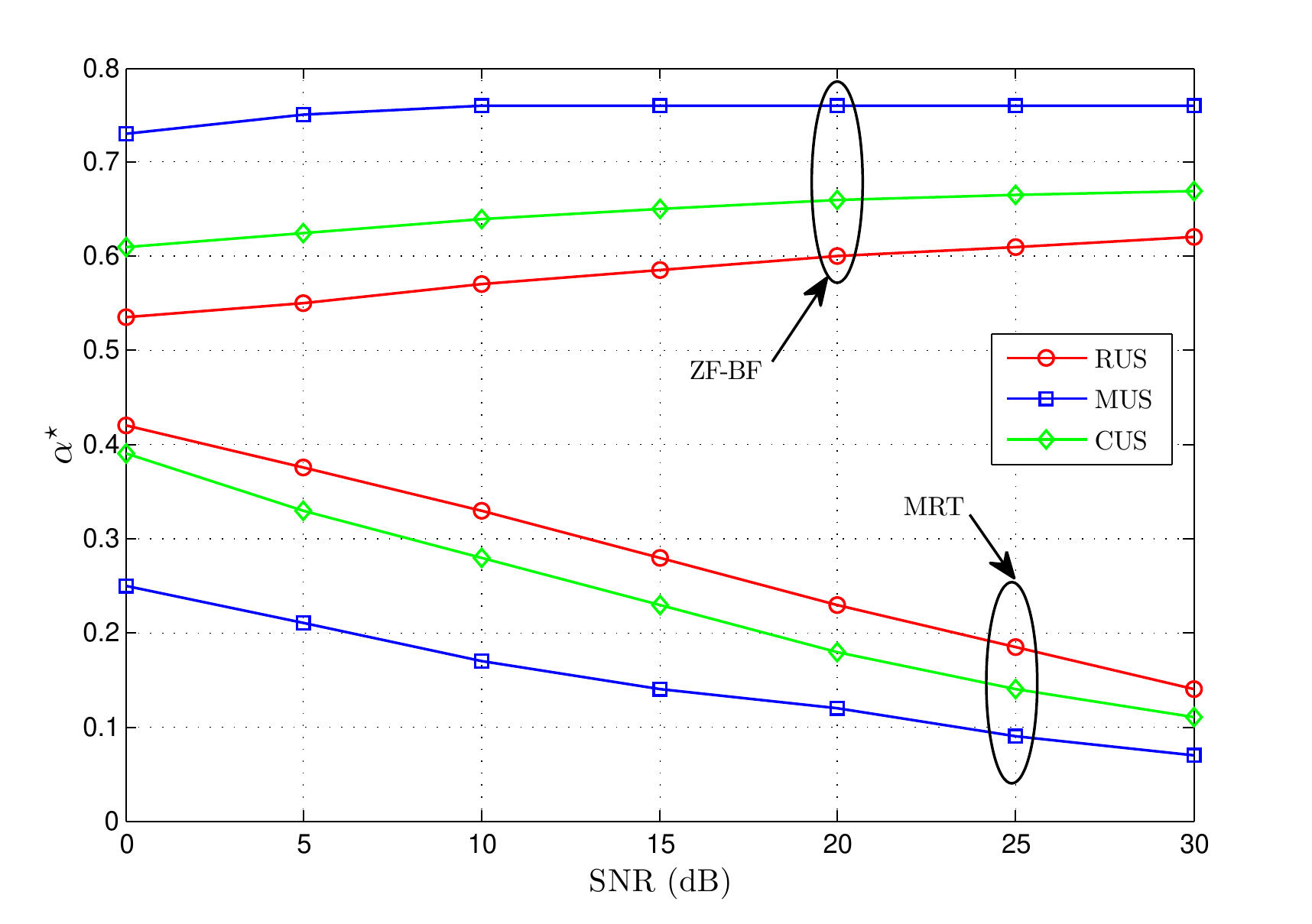}
	\vspace{-0.8cm} 
	\caption{$\alpha^\star$ vs. SNR, where $K=10$.}
	\label{fig:alpha_star_SNR_K10}
	\includegraphics[width=3.45in]{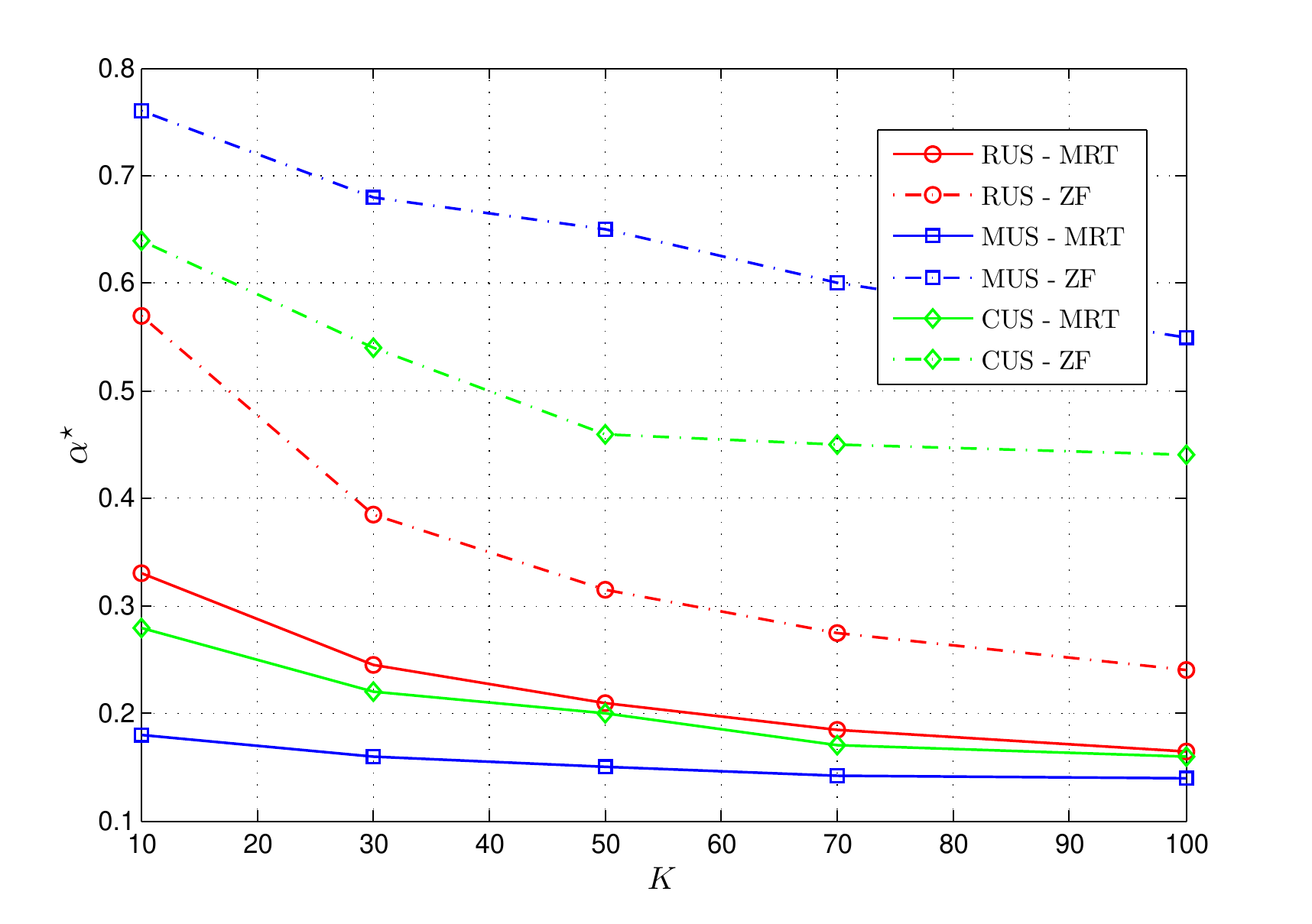}
	\vspace{-0.8cm} 
	\caption{$\alpha^\star$ vs. $K$, where SNR = 10dB.}
	\label{fig:alpha_star_K}
\end{figure}


Figure~\ref{fig:SUS_K10} shows the ergodic sum-rate vs. $\alpha$ for the MUS scheme for different transmit SNR levels. Similar figures for the RUS and CUS scheme can also be obtained but are omitted due to space limitations. The number of users is set at $K=10$. The red solid lines represent the theoretical ergodic sum-rate of the MUS scheme using MRT; and the blue dashed lines represent the theoretical ergodic sum-rate of the MUS scheme using ZF-BF. The points marked with $\Circle$ and $\Diamond$ represent the average sum-rate of MRT and ZF-BF obtained by simulation, respectively. As can be observed from the figure, the curves generated from the theoretical ergodic sum-rate expressions match well with the simulation results. As expected, the ergodic sum-rate changes with $\alpha$. We can see that for ZF-BF, the ergodic sum-rate is roughly constant for a range of $\alpha$ between $0.6$ to $0.8$. On the other hand, MRT is more sensitive to the choice of $\alpha$, as it directly controls the amount of interference experienced by the users. 
The optimal $\alpha$ that maximizes the ergodic sum-rate (denoted by $\alpha^\star$) behaves differently for MRT and ZF-BF when the SNR is increased. This observation will be discussed further with regards to Fig.~\ref{fig:alpha_star_SNR_K10}.


Figure~\ref{fig:alpha_star_SNR_K10} shows the behaviour of $\alpha^\star$ as the transmit SNR changes. As can be observed from the figure, $\alpha^\star$ is strictly decreasing with transmit SNR for MRT. In contrast, $\alpha^\star$ slightly increases with increasing SNR for ZF-BF. This is because for MRT, the system becomes interference-limited at high SNR, hence we need to reduce $\alpha$ to decrease interference. On the other hand, the interference is always zero for ZF-BF, hence $\alpha$ is not highly sensitive to the change in SNR.

Figure~\ref{fig:alpha_star_K} shows the behaviour of $\alpha^\star$ as the number of users $K$ changes. Intuitively, as $K$ increases, we can afford to be more demanding on the degree of orthogonality while keeping a good number of semi-orthogonal users, \emph{i.e.}, the expected number of semi-orthogonal users is $(K-1) \alpha^2$, and when $K$ increases, we can afford to drop $\alpha$ (and thus reduce interference) without sacrificing the expected number of semi-orthogonal users. 
This figure explores the trade-off between multi-user diversity gain (resulting from having more semi-orthogonal users by maintaining or increasing $\alpha$), and directional diversity gain (resulting from having less interference by decreasing $\alpha$). As can be observed from the figure, $\alpha^\star$ decreases with $K$. So when there are more users in the system, we should trade multi-user diversity gain for interference reduction (or directional diversity gain).

\begin{figure}[t]
\centering
\includegraphics[width=3.42in]{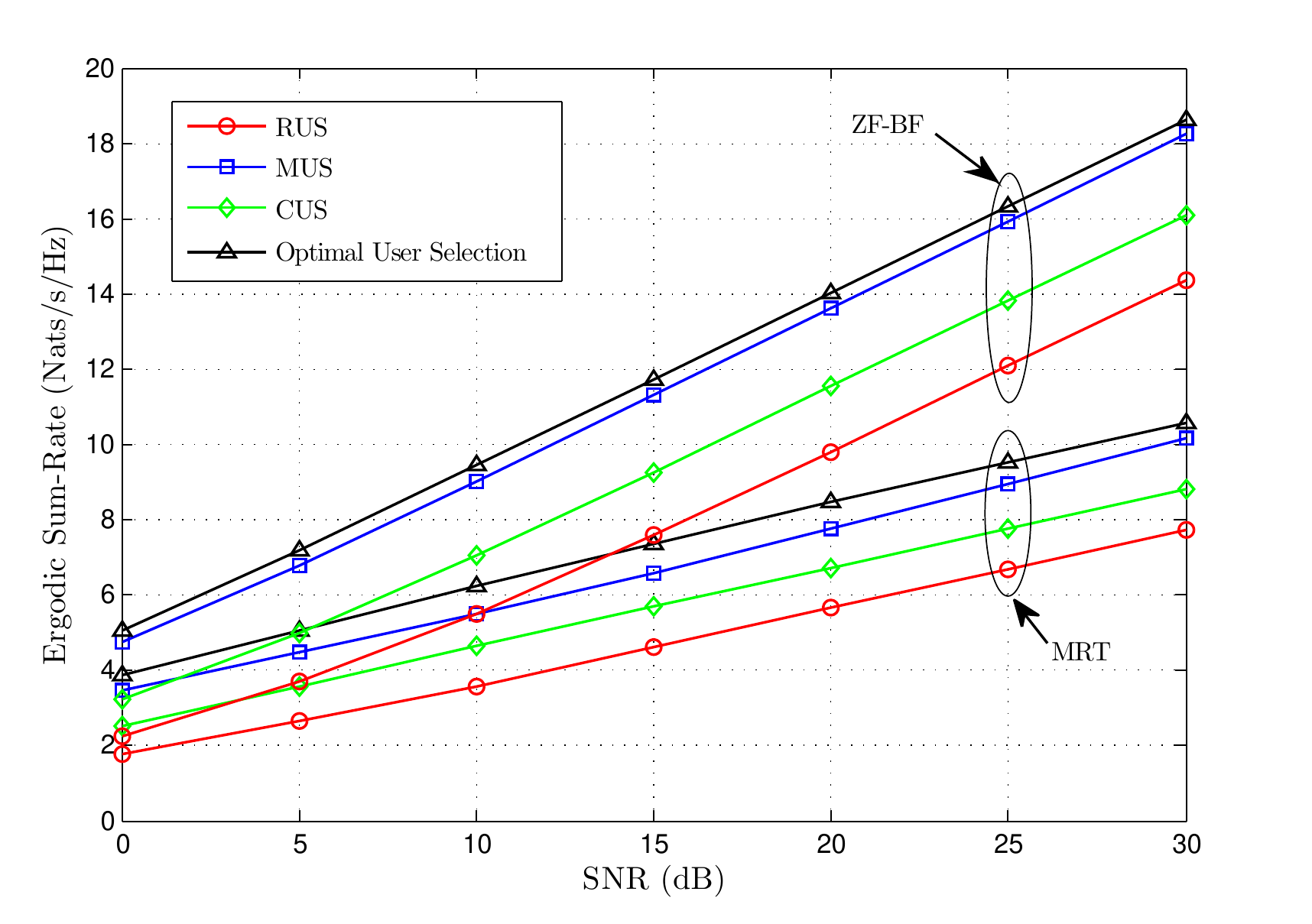}
\vspace{-0.4cm}
\caption{Ergodic Sum-Rate vs. SNR, where $\alpha=\alpha^\star$ and $K=10$.}
\label{fig:SumRate_all3_optimized_alpha}
\includegraphics[width=3.42in]{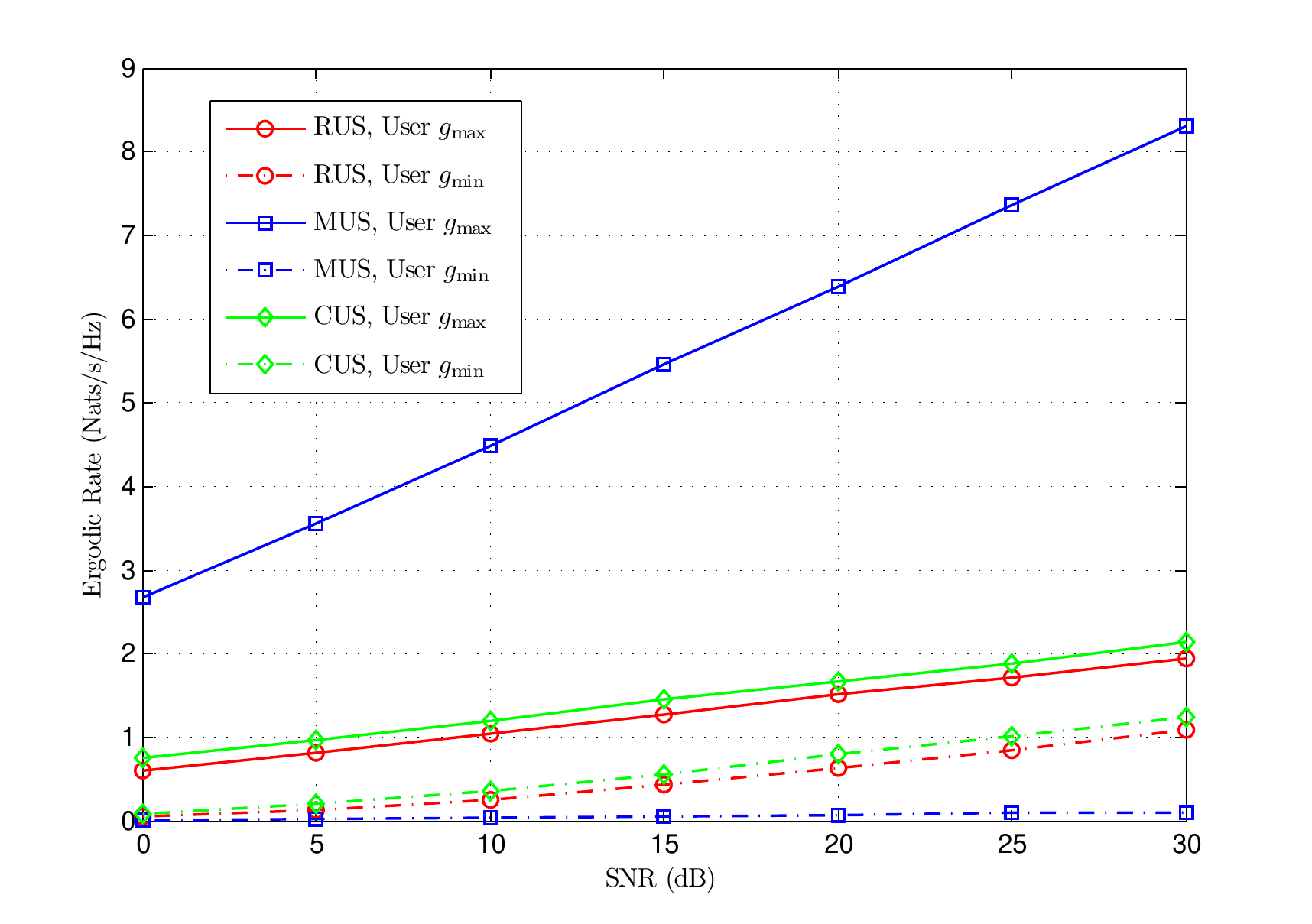}
\vspace{-0.4cm}
\caption{Ergodic Rate Comparison between user with $g_{\max}$ and $g_{\min}$ vs. SNR, $K=10$, $\alpha=\alpha^\star$.}
\label{fig:fairness_comparison}
\end{figure}

Figure~\ref{fig:SumRate_all3_optimized_alpha} shows the ergodic sum-rate vs. SNR when $\alpha$ is set at $\alpha^\star$ for each scheduling scheme and beamforming method. The black line marked with $\Delta$ represents the optimal scheduled user set via exhaustive search. As expected, the MUS scheme achieves the highest ergodic sum-rate (which is very close to the sum-rate achieved with the optimal user set), while the RUS scheme achieves the lowest ergodic sum-rate, for both MRT and ZF-BF. This is because the MUS scheme utilizes the multi-user diversity gains fully. 

Figure~\ref{fig:fairness_comparison} compares the ergodic rates of the user with the largest PL value (denoted by $g_{\max}$) and the user with the smallest PL value (denoted by $g_{\min}$), as the transmit SNR varies. The number of users $K$ is set at $10$, the BS is assumed to use ZF-BF, and $\alpha$ is set at $\alpha^\star$. The RUS and the CUS schemes provide fairness as each user has an equal probability to be selected. As a consequence, we can observe that the user with the smallest PL value will achieve an ergodic rate comparable to the user with the largest PL value. However, this is not the case for the MUS scheme, as the ergodic rate of the user with the largest PL value is much higher compared to the user with the smallest PL value. Intuitively, the user with the smallest PL value is almost never selected as $\pi(1)$, and the most likely scenario for it to be selected as $\pi(2)$ is by being the only user semi-orthogonal to $\pi(1)$, hence resulting in a very small rate.

\begin{figure}[t]
	\centering
		\includegraphics[width=3.4in]{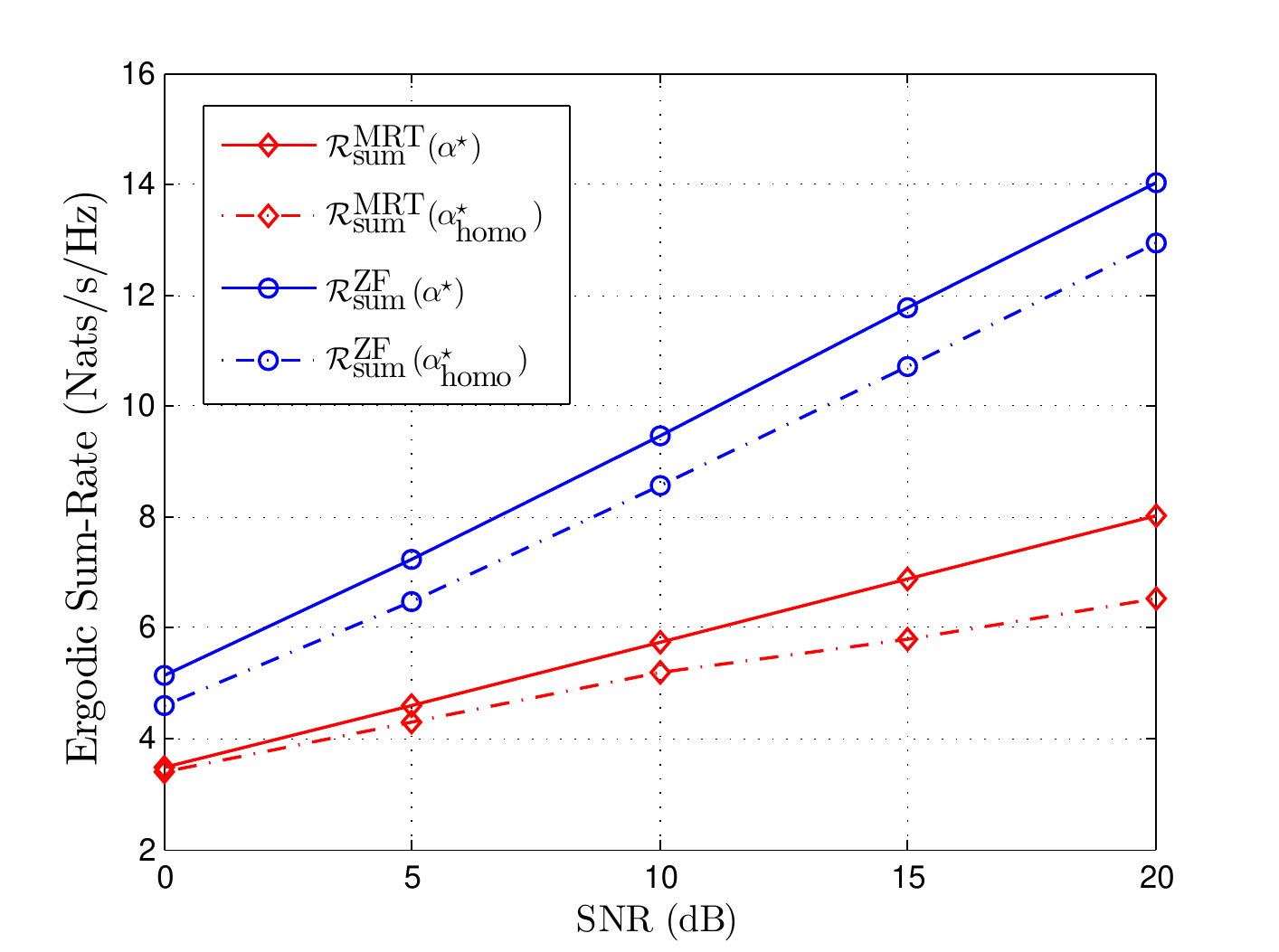}
	\caption{Ergodic Sum-Rate vs. SNR with $\alpha^\star$ and $\alpha^\star_{\mbox{\tiny{homo}}}$.}
	\label{fig:new_fig}
\end{figure}
Figure~\ref{fig:new_fig} compares the ergodic sum-rate using $\alpha^\star$ and $\alpha^\star_{\mbox{\tiny{homo}}}$ for different SNR value; $\alpha^\star$ is the optimal $\alpha$ that maximizes the ergodic sum-rate for the specific realization of the users, and $\alpha^\star_{\mbox{\tiny{homo}}}$ is the sum-rate maximizing $\alpha$ when the PL values are neglected (or all $g$'s are set at one). We use both $\alpha^\star$ and $\alpha^\star_{\mbox{\tiny{homo}}}$ to calculate the ergodic sum-rate for the aforementioned realization of the users where each user had its own path loss value. To this end, the solid lines represent the ergodic sum-rate achieved with $\alpha^\star$, and the dotted lines represent the ergodic sum-rate achieved with $\alpha^\star_{\mbox{\tiny{homo}}}$. As expected, the sum-rate achieved with $\alpha^\star_{\mbox{\tiny{homo}}}$ is lower than the sum-rate achieved with $\alpha_\star$ for both MRT and ZF-BF, and the rate-loss is greater as we increase the SNR. This indicates the benefit of carrying out the analysis for heterogeneous users.


\section{Conclusions}\label{sec:ch3_conclusion}
In this paper, we focused on the ergodic downlink sum-rate performance of a system consisting of a set of heterogeneous users. We studied three user scheduling schemes to group near-orthogonal users for simultaneous transmission, namely, the Random User Selection (RUS) scheme, the Max-Gain User Selection (MUS) scheme and the CDF-Based User Selection (CUS) scheme. The RUS scheme focuses on simplicity and scheduling fairness, and does not exploit the multi-user diversity gain; the MUS scheme fully exploits multi-user diversity gain and does not achieve scheduling fairness; finally, the CUS scheme partially exploits multi-user diversity while guaranteeing scheduling fairness. For each scheduling scheme, we considered maximum-ratio transmission (MRT) and zero-forcing beamforming (ZF-BF) for data transmission. 

In all of the scheduling schemes considered in the paper, there is a system parameter $\alpha$ that controls the degrees of orthogonality of channel directions between co-scheduled users, and the system performance is sensitive to $\alpha$. We obtained analytical expressions for the ergodic downlink sum-rate considering each scheduling scheme, for both MRT and ZF-BF. We focused on the problem of finding $\alpha^\star$, which is the optimal $\alpha$ that maximizes the ergodic downlink sum-rate of the system. Numerical results indicated the following behaviours of $\alpha^\star$: 1) $\alpha^\star$ decreased with the number of users in the system; 2)  when using MRT, $\alpha^\star$ decreased with transmit SNR for all scheduling schemes; and 3) when using ZF-BF, $\alpha^\star$ slightly increased with transmit SNR.


\appendices
\section{Proofs for the RUS Scheme, MU-Mode}
\subsection{MRT - Proof of Lemma \ref{lem:RUS_MUMode_MRT}} \label{app:proof_of_lemma_RUS_MUMode_MRT} 
Let $X=\frac{\hat{\sigma}^2_{k}}{\|\mathbf{h}_{k}\|^2}$. The random variable $X$ is inverse-Gamma distributed with parameters $2$ and $\hat{\sigma}^2_{k}$, \emph{i.e.}, $F_X(x) = \Gamma(2,\hat{\sigma}^2_{k} x)$ where $\Gamma(\cdot,\cdot)$ is the upper incomplete Gamma function. Let $Z=X+Y^{\mbox{\tiny{MRT}}}$ where $Y^{\mbox{\tiny{MRT}}} = |\tilde{\mathbf{h}}_i^{\textrm{H}} \tilde{\mathbf{h}}_j|^2$. The PDF of $Z$ can be obtained by convolving the PDFs of $X$ and $Y^{\mbox{\tiny{MRT}}}$, \emph{i.e.}, 
$f_Z(x) = \int_0^\infty f_X(x-t) f_{Y^{\mbox{\tiny{MRT}}}}(t) dt$. 
The integrand is zero unless $(x-t)\geq 0$. So, if $0 \leq x \leq \alpha^2$, we have 
	$f_Z(x) = \frac{1}{\alpha^2} \int_0^x f_X(x-t) dt$, 
and if $x\geq \alpha^2$, we have 
	$f_Z(x) = \frac{1}{\alpha^2}\int_0^{\alpha^2} f_X(x-t) dt$. 
Evaluating the integrals and setting $f_{\mbox{\tiny{SINR}}_{k}}(x) = \frac{f_Z(\frac{1}{x})}{x^2}$ yields the SINR distribution of each selected user as
\begin{align*} 
	f_{\mbox{\tiny{SINR}}_{k}}(x) {=} \left\{ \begin{array}{l} \frac{1}{\alpha^2 x^2} \left[\Gamma(2,\hat{\sigma}^2_{k} x) - \Gamma\left(2, \frac{\hat{\sigma}^2_{k}}{\frac{1}{x}-\alpha^2} \right) \right]  0 \leq x \leq \frac{1}{\alpha^2} \\ 
	\frac{1}{\alpha^2 x^2}\Gamma(2, \hat{\sigma}^2_{k} x) ~~~~~~~~~~~~~~~~~~~~~ \mbox{otherwise} \end{array} \right. ,
\end{align*}
where $\hat{\sigma}^2_{k} = \sigma^2/g_{k}$. By using the above SINR distribution, the ergodic rate of user $k$ can be written as
\begin{align}
& \mathbb{E}\left[\log(1+\mbox{SINR}_k)\right] = \int_{0}^{\infty} \log(1+x) f_{\mbox{\tiny{SINR}}_{k}}(x) dx \nonumber \\
&=\frac{1}{\alpha^2} \left[\int_0^\infty \frac{\log(1+x)}{x^2} \Gamma(2,\hat{\sigma}^2_k x) dx  \right. \nonumber \\ & \hspace{2cm} \left.- \int_0^{\frac{1}{\alpha^2}} \frac{\log(1+x)}{x^2} \Gamma\left(2, \frac{\hat{\sigma}^2_{k}}{\frac{1}{x}-\alpha^2}\right) dx \right]  \nonumber \\
&= G\left(\frac{\hat{\sigma}^2_{k}}{\alpha^2}\right) + \frac{1}{\alpha^2} G(\hat{\sigma}^2_{k}) -  \frac{\alpha^2+1}{\alpha^2} G\left(\frac{\hat{\sigma}^2_{k}}{1+\alpha^2}\right). \label{eq:conditional_rate_mumode_rus_mrt}
\end{align}
In the RUS scheme, each user $k\in\{1,\ldots,K\}$ is equally likely to be selected as $\pi(1)$, \emph{i.e.}, $\Pr\{\pi(1) = k\} = \frac{1}{K}$. Averaging (\ref{eq:conditional_rate_mumode_rus_mrt}) over $k$ yields $\mathcal{R}_{M1}^{\mbox{\tiny{MRT}}}(\alpha)$. 

Let $\mathcal{X}$ denote the set of users semi-orthogonal to $\pi(1)$. For the RUS scheme, each user in $\mathcal{X}$ is equally likely to be selected as $\pi(2)$. Therefore, we have
\begin{align} \label{eq:new2}
	\Pr\{\pi(2) = k | \mbox{MU-Mode}, k\in\mathcal{X}, |\mathcal{X}|=m\} = \frac{1}{m}. 
\end{align}
The conditional probability of $k\in\mathcal{X}$ given $|\mathcal{X}|=m$ can be calculated as
\begin{align*}
 \Pr\{k \in \mathcal{X} | |\mathcal{X}|=m\} = \frac{\binom{K-1}{m} - \binom{K-2}{m}}{\binom{K-1}{m}} = \frac{\binom{K-2}{m-1}}{\binom{K-1}{m}} = \frac{m}{K-1},
\end{align*}
where the term $\binom{K-1}{m}$ represents the number of subsets of cardinality $m$, and $\binom{K-2}{m}$ represents the number of subsets that does not contain user $k$. The joint probability of $k\in\mathcal{X}$ and $|\mathcal{X}|=m$ is given by
\begin{align*}
	\Pr\{k\in\mathcal{X}, |\mathcal{X}|=m \} = \frac{m}{K-1} \Pr\{|\mathcal{X}|=m\}. 
\end{align*}
Averaging (\ref{eq:new2}) over $k\in\mathcal{X}, |\mathcal{X}|=m$ yields
\begin{align}
	\Pr\{\pi(2) = k |& \mbox{MU-Mode}\} \nonumber \\
	&= \sum_{m=1}^{K-1} \left(\frac{1}{m}\right) \left(\frac{m}{K-1}\right) \Pr\{|\mathcal{X}|=m\} \nonumber \\
	 							&=\frac{1}{K-1} \sum_{m=1}^{K-1} \Pr\{|\mathcal{X}|=m\} \nonumber \\
								&= \frac{1}{K-1}. \nonumber 
\end{align}
Therefore, each of the remaining $K-1$ user is equally likely to be selected as $\pi(2)$, given that we are operating in the MU-Mode. The conditional rate of $\pi(2)$, given $\pi(1)$ under MU-Mode, is
\begin{align} \label{eq:appendix_A_conditional_rate}
	\mathcal{R}_{M2}^{\mbox{\tiny{MRT}}}(\alpha|\pi(1)) &= \frac{1}{K-1} \sum_{\substack{k\neq \pi(1) \\ k=1}}^K \left[ G\left(\frac{\sigma^2}{g_k\alpha^2}\right) + \frac{1}{\alpha^2} G\left(\frac{\sigma^2}{g_k}\right) \right. \nonumber \\
	& \hspace{2.5cm} \left. - \frac{\alpha^2+1}{\alpha^2} G\left(\frac{\sigma^2}{g_k(1+\alpha^2)}\right) \right].
\end{align}
Averaging (\ref{eq:appendix_A_conditional_rate}) over $\pi(1)$ yields
\begin{align*}
	\mathcal{R}_{M2}^{\mbox{\tiny{MRT}}}(\alpha) &= \frac{1}{K} \sum_{i=1}^K \frac{1}{K-1} \sum_{\substack{k\neq i \\ k=1}}^K \left[ G\left(\frac{\sigma^2}{g_k\alpha^2}\right) + \frac{1}{\alpha^2} G\left(\frac{\sigma^2}{g_k}\right) \right. \nonumber \\ & \hspace{3.5cm} \left. - \frac{\alpha^2+1}{\alpha^2} G\left(\frac{\sigma^2}{g_k(1+\alpha^2)}\right) \right] \\
						&= \frac{1}{K} \sum_{k=1}^K \left[ G\left(\frac{\sigma^2}{g_k\alpha^2}\right) + \frac{1}{\alpha^2} G\left(\frac{\sigma^2}{g_k}\right) \right. \nonumber \\ & \hspace{3cm} \left. - \frac{\alpha^2+1}{\alpha^2} G\left(\frac{\sigma^2}{g_k(1+\alpha^2)}\right) \right],
\end{align*}
which completes the proof.

\subsection{ZF - Proof of Lemma \ref{lem:RUS_MUMode_ZF}} \label{app:proof_of_lemma_RUS_MUMode_ZF} 
When the BS operates under MU-Mode using ZF-BF, the inter-user interference would be zero. The distribution of $\mbox{SINR}_{k}$ is given by the product distribution of $S_{k}$ and $Y^{\mbox{\tiny{ZF}}}$, where $Y^{\mbox{\tiny{ZF}}} = |\tilde{\mathbf{h}}_i^{\textrm{H}} \mathbf{w}_i|^2$, \emph{i.e.}, 
$f_{\mbox{\tiny{SINR}}_{k}}(x)= \int_{0}^{\infty} f_{S_{k}}(t) f_{Y^{\mbox{\tiny{ZF}}}}\left(\frac{x}{t}\right) \frac{1}{t} dt$.
The integrand is zero unless $1-\alpha^2 \leq \frac{x}{t}\leq 1$, so we have 
\begin{align*}
f_{\mbox{\tiny{SINR}}_{k}}(x) &= \frac{\hat{\sigma}_{k}^4}{\alpha^2} \int_{x}^{\frac{x}{1-\alpha^2}} \exp(-\hat{\sigma}_{k}^2 t) dt \\
	&= \frac{\hat{\sigma}^2_{k}}{\alpha^2} \left[\exp(-\hat{\sigma}^2_{k} x) - \exp\left(-\frac{\hat{\sigma}^2_{k}x}{1-\alpha^2}\right) \right],
\end{align*}
where $\hat{\sigma}^2_{k} = \sigma^2/g_{k}$. By using the above SINR distribution, the ergodic rate of user $k$ can be written as
\begin{align}
\mathbb{E} [\log(1+\mbox{SINR}_k)] &= \int_0^\infty \log(1+x) f_{\mbox{\tiny{SINR}}_{k}}(x) dx \nonumber\\
&= \frac{1-\alpha^2}{\alpha^2} G\left(\frac{\hat{\sigma}^2_{k}}{1-\alpha^2}\right) - \frac{G(\hat{\sigma}^2_{k})}{\alpha^2}. \label{eq:conditional_rate_mumode_rus_zf}
\end{align}
As we have seen in the proof of Lemma \ref{lem:RUS_MUMode_MRT}, each user is equally likely to be selected as $\pi(1)$ and $\pi(2)$. Therefore, averaging (\ref{eq:conditional_rate_mumode_rus_zf}) over $k$ completes the proof.

\section{Proofs for the MUS Scheme, MU-Mode}
\subsection{First Selected User - Proof of Lemma \ref{lem:SUS_MRT_MUmode_1stuser}}\label{app:proof_of_lemma_SUS_MRT_MUmode_1stuser} 
For the MUS scheme, the CDF of $S_{\pi(1)}$ is given by (\ref{eq:SUS_SNR_User1}). The PDF of $S_{\pi(1)}$ can be written as
\begin{align*}
	f_{S_{\pi(1)}}(x) = \sum_{k=1}^K \frac{\sigma^4}{g_k^2} x \exp\left(-\frac{\sigma^2}{g_k}x\right) \prod_{\substack{j=1 \\ j\neq k}}^K \gamma\left(2, \frac{\sigma^2}{g_k}x\right).
\end{align*}
Therefore, the ergodic rate of the first selected user using MRT can be calculated as
\begin{align*}
&\mathcal{R}^{\mbox{\tiny{MRT}}}_{M1}(\alpha) 
	{=} \int_0^\infty  \hspace{-0.2cm} \int_0^{\alpha^2} \hspace{-0.2cm} \log(1+(x^{-1}+y)^{-1}) dF_{Y^{\mbox{\tiny{MRT}}}}(y) dF_{S_{\pi(1)}}(x)\nonumber \\
	&{=} \frac{\sigma^4}{\alpha^2} \sum_{k=1}^K \frac{1}{g_k^2} \int_0^\infty \Upsilon^{\mbox{\tiny{MRT}}}(x) x\exp\left(-\frac{\sigma^2}{g_k} x\right) \prod_{\substack{j=1\\j\neq k}}^{K} \gamma\left(2,\frac{\sigma^2}{g_j}x\right) dx,
\end{align*}
and the ergodic rate of the first selected user using ZF-BF can be calculated as
\begin{align*}
&\mathcal{R}_{M1}^{\mbox{\tiny{ZF}}}(\alpha) = \int_0^\infty \int_{1-\alpha^2}^1 \hspace{-0.4cm} \log(1+xy) dF_{Y^{\mbox{\tiny{ZF}}}}(y) dF_{S_{\pi(1)}}(x) \nonumber \\
&= \frac{\sigma^4}{\alpha^2} \sum_{k=1}^K \frac{1}{g_k^2} \int_0^\infty \Upsilon^{\mbox{\tiny{ZF}}}(x) x\exp\left(-\frac{\sigma^2}{g_k}x\right) \prod_{\substack{j=1 \\ j\neq k}}^{K} \gamma\left(2,\frac{\sigma^2}{g_j}x\right) dx,
\end{align*}
which completes the proof.

\subsection{Second Selected User - Proof of Lemma \ref{lem:SUS_MRT_MUmode_2nduser}}\label{app:proof_of_lemma_SUS_MRT_MUmode_2nduser} 
We first order the users according to $S_{(1)} > S_{(2)} > \cdots >S_{(K)}$. The PDF of $S_{\pi(2)}$ under MU-Mode can be calculated as
\begin{align} \label{eq:pdf_of_SNR_user2_SUS_MUMode}
	f_{S_{\pi(2)}}(x|\mbox{\small{MU-Mode}}) = \sum_{i=2}^K \Pr\{ \pi(2) = (i) | \mbox{\small{MU-Mode}} \} f_{S_{(i)}}(x),
\end{align}
where in~\cite{Lu2009}, $\Pr\{ \pi(2) = (i) | \mbox{\small{MU-Mode}} \}$ is given by
\begin{align} \label{eq:probability_of_ith_rank}
	\Pr\{ \pi(2) = (i) | \mbox{\small{MU-Mode}} \} = \frac{\alpha^2 (1-\alpha^2)^{i-2}}{1-\lambda},
\end{align}
and $f_{S_{(i)}}(x)$ is the PDF of the $i$th order statistic out of $K$ i.n.i.d. RVs. According to~\cite{non_iid_order_stats},
\begin{align*} 
&f_{S_{(i)}}(x) = \frac{1}{(i-1)! (K-i)!} \times \\ & \begin{matrix}  \hspace{-0.7cm} \left. \begin{array}{c} \\ \\ \\ \\ \\ \\ \\ \end{array} \right.^+
\underbrace{\begin{vmatrix} 
					F_{S_1}(x) & \cdots & F_{S_K}(x) \\
					\cdots & \cdots & \cdots \\
					F_{S_1}(x) & \cdots & F_{S_K}(x) \\
					f_{S_1}(x) & \cdots & f_{S_K}(x) \\
					1-F_{S_1}(x) & \cdots & 1-F_{S_K}(x) \\
						\cdots & \cdots & \cdots \\
					1-F_{S_1}(x) & \cdots & 1-F_{S_K}(x) 
				\end{vmatrix}^+}_{\mathbf{A}_{(i)}}
			\begin{aligned}
			&\left.\begin{matrix} \\ \\ \\ \end{matrix} \right\} %
			K-i \mbox{ identical rows} \\
			&\left.\begin{matrix} \\ \end{matrix} \right.\\%
			&\left. \begin{matrix} \\ \\ \\ \end{matrix} \right\}%
			i-1 \mbox{ identical rows}\\
		\end{aligned}
	\end{matrix}
\end{align*}
where ${}^+|\mathbf{A}_{(i)}|^+$ denote the \emph{permanent} of the $K$-by-$K$ matrix $\mathbf{A}_{(i)}$. To this end, since the $(K-i+1)$th row of $\mathbf{A}_{(i)}$ is an unique row, ${}^+|\mathbf{A}_{(i)}|^+$ can be further simplified, and we have
\begin{align*}
	{}^+|\mathbf{A}_{(i)}|^+ = \sum_{k=1}^K f_{S_k}(x)~~{}^+|\mathbf{A}_{K-i+1,k}|^+,
\end{align*}
where $\mathbf{A}_{K-i+1,k}$ is a $(K-1)$-by-$(K-1)$ matrix obtained by removing the $(K-i+1)$th row and $k$th column of $\mathbf{A}_{(i)}$. 
${}^+|\mathbf{A}_{K-i+1,k}|^+$ can be simplified as
\begin{multline*} 
\hspace{-0.3cm}	{}^+|\mathbf{A}_{K-i+1,k}|^+ = (i-1)! (K-i)!~\times \\ \sum_{m=1}^{\binom{K-1}{K-i}} \left[ \prod_{j=1}^{K-i} F_{S_{\mathbf{C}_{(i)}^k(m,j)}}(x) \prod_{j=1}^{i-1} \Big[1-F_{S_{\overline{\mathbf{C}}_{(i)}^k(m,j)}}(x)\Big] \right].
\end{multline*}
Therefore, the PDF of $S_{(i)}$ can be written as
\begin{align} \label{eq:SUS_ordered_stats_2}
&f_{S_{(i)}}(x) = \nonumber \\
	& \sum_{k=1}^K f_{S_k}(x)  \sum_{n=1}^{\binom{K-1}{K-i}} \prod_{j=1}^{K-i} F_{S_{\mathbf{C}_{(i)}^k(n,j)}}(x) \prod_{j=1}^{i-1} \Big[1-F_{S_{\overline{\mathbf{C}}_{(i)}^k(n,j)}}(x)\Big]. 
\end{align}
Substituting (\ref{eq:probability_of_ith_rank}) and (\ref{eq:SUS_ordered_stats_2}) into (\ref{eq:pdf_of_SNR_user2_SUS_MUMode}) yields
\begin{multline} \label{eq:pdf_of_SNR_user2_SUS_MUMode_final}
	f_{S_{\pi(2)}}(x|\mbox{\small{MU-Mode}}) = \sum_{i=2}^K \frac{\alpha^2 (1-\alpha^2)^{i-2}}{1-\lambda}  
											 \sum_{k=1}^{K} f_{S_k}(x) \\ \sum_{n=1}^{\binom{K-1}{K-i}}  \prod_{j=1}^{K-i} F_{S_{\mathbf{C}_{(i)}^k(n,j)}}(x) \prod_{j=1}^{i-1} \Big[1-F_{S_{\overline{\mathbf{C}}_{(i)}^k(n,j)}}(x)\Big].
\end{multline}
By using (\ref{eq:pdf_of_SNR_user2_SUS_MUMode_final}), the ergodic rate of the second selected user in MU-Mode using MRT can be written as
\begin{align*}
	&\mathcal{R}^{\mbox{\tiny{MRT}}}_{M2} (\alpha) \\
	&{=} \hspace{-0.1cm} \int_0^\infty \hspace{-0.2cm} \int_0^{\alpha^2} \hspace{-0.2cm} \log(1+(x^{-1}+y)^{-1}) dF_{Y^{\mbox{\tiny{MRT}}}}(y) f_{S_{\pi(2)}}(x|\mbox{\small{MU-Mode}})  dx \nonumber \\
	&= \sum_{i=2}^K   \frac{(1-\alpha^2)^{i-2}}{1-\lambda}  \int_0^\infty \Upsilon^{\mbox{\tiny{MRT}}}(x) \sum_{k=1}^{K} \frac{\sigma^4}{g_k^2} x \exp\left(-\frac{\sigma^2}{g_k}x\right) \nonumber \\ & \hspace{0.5cm} \sum_{m=1}^{\binom{K-1}{K-i}} \prod_{j=1}^{K-i} \gamma\left(2, \frac{\sigma^2}{g_{\mathbf{C}_{(i)}^k(m,j)}}x\right) \prod_{j=1}^{i-1} \Gamma\left(2, \frac{\sigma^2}{g_{\overline{\mathbf{C}}_{(i)}^k(m,j)}} x\right) dx,
\end{align*}
and the ergodic rate of the second selected user in MU-Mode using ZF-BF can be calculated as
\begin{align*}
 &\mathcal{R}_{M2}^{\mbox{\tiny{ZF}}} (\alpha) {=} \int_0^\infty \hspace{-0.2cm} \int_{1-\alpha^2}^1 \hspace{-0.2cm} \log(1+xy) dF_{Y^{\mbox{\tiny{ZF}}}}(y) f_{S_{\pi(2)}}(x|\mbox{\small{MU-Mode}})  dx \nonumber \\
	&= \sum_{i=2}^K \frac{  (1-\alpha^2)^{i-2}}{1-\lambda}  \int_0^\infty \Upsilon^{\mbox{\tiny{ZF}}}(x) \sum_{l=1}^{K} \frac{\sigma^4}{g_l^2} x \exp\left(-\frac{\sigma^2}{g_l}x\right) \nonumber \\ & \hspace{0.5cm} \sum_{m=1}^{\binom{K-1}{K-i}} \prod_{j=1}^{K-i} \gamma\left(2, \frac{\sigma^2}{g_{\mathbf{C}_{(i)}^k(m,j)}}x\right) \prod_{j=1}^{i-1} \Gamma\left(2, \frac{\sigma^2}{g_{\overline{\mathbf{C}}_{(i)}^k(m,j)}} x\right) dx,
\end{align*}
which completes the proof.

\section{Proofs for the CUS Scheme, MU-Mode}
\subsection{First Selected User - Proof of Lemma~\ref{lem:CUS_User1_MRT}}\label{app:proof_of_lemma_CUS_User1_MRT} 
The CDF of $S_{\pi(1)}$ for the CUS scheme is given by (\ref{eq:CDF_of_SNR_1stuser_CUS}). Therefore, the ergodic rate of the first selected user under MU-Mode using MRT can be calculated as
\begin{align*}
&\mathcal{R}_{M1}^{\mbox{\tiny{MRT}}} (\alpha) {=} \int_0^\infty \hspace{-0.2cm} \int_0^{\alpha^2} \hspace{-0.2cm} \log(1+(x^{-1}+y)^{-1}) dF_{Y^{\mbox{\tiny{MRT}}}}(y) dF_{S_{\pi(1)}}(x) \nonumber \\
&= \frac{1}{\alpha^2} \sum_{k=1}^{K}  \int_0^\infty  \hspace{-0.2cm}\Upsilon^{\mbox{\tiny{MRT}}}(x) \left[\gamma\left(2, \frac{\sigma^2}{g_k} x\right)\right]^{K-1} \hspace{-0.5cm} x\exp\left(-\frac{\sigma^2}{g_k}x\right) \frac{\sigma^4}{g_k^2} dx,
\end{align*}
and the ergodic rate of the first selected user under MU-Mode using ZF-BF can be calculated as
\begin{align*}
&\mathcal{R}_{M1}^{\mbox{\tiny{ZF}}} (\alpha) {=} \int_0^\infty \hspace{-0.2cm} \int_{1-\alpha^2}^1 \hspace{-0.2cm} \log(1+xy) dF_{Y^{\mbox{\tiny{ZF}}}}(y) dF_{S_{\pi(1)}}(x) \nonumber \\
&= \frac{1}{\alpha^2} \sum_{k=1}^{K}  \int_0^\infty \hspace{-0.2cm} \Upsilon^{\mbox{\tiny{ZF}}}(x) \left[\gamma\left(2, \frac{\sigma^2}{g_k} x\right)\right]^{K-1} \hspace{-0.5cm} x\exp\left(-\frac{\sigma^2}{g_k}x\right) \frac{\sigma^4}{g_k^2} dx,
\end{align*}
which completes the proof.

\subsection{Second Selected User - Proof of Lemma~\ref{lem:CUS_MRT_User2}}\label{app:proof_of_lemma_CUS_User2_MRT} 
We will first derive the conditional distribution of $S_{\pi(2)}$, given $\pi(1), p_{\pi(1)}$, and $\mathcal{X}$. Then, we will average over $p_{\pi(1)}$, $\mathcal{X}$, and $\pi(1)$, respectively. To this end, 
\begin{align*}
&F_{S_{\pi(2)}}(x|\pi(1), p_{\pi(1)}, \mathcal{X}) =  \Pr\{S_{\pi(2)}\leq x| \pi(1), p_{\pi(1)}, \mathcal{X}\} \nonumber \\
	&= \sum_{k\in\mathcal{X}} \Pr\{p_k \leq F_{S_{k}}(x), p_k \leq p_{\pi(1)}, p_j \leq p_k, \nonumber \\ & \hspace{5cm} \forall j\in\mathcal{X}, j \neq k | p_{\pi(1)}, \mathcal{X}\} \nonumber \\
	&= \sum_{k\in\mathcal{X}} \int_0^{\min\left(F_{S_k}(x),~p_{\pi(1)}\right)} \hspace{-2.4cm} \Pr\{p_k=t\} \Pr\{p_j\leq t, \forall j\in\mathcal{X}, j \neq k | \mathcal{X}, p_k=t\} dt.
\end{align*}
Given $p_k \leq p_{\pi(1)}$ where $k\in\mathcal{X}$, $p_k$ becomes uniformly distributed over $[0,p_{\pi(1)}]$, \emph{i.e.}, $f_{p_k}(x|p_k \leq p_{\pi(1)}) = \frac{f_{p_k}(x)}{F_{p_k}(p_{\pi(1)})} = \frac{1}{p_{\pi(1)}}$. 
Therefore, we have $\Pr\{p_k=t\} = \frac{1}{p_{\pi(1)}}$ and 
\begin{align*}
\Pr \{p_j\leq t, \forall j\in \mathcal{X}, j\neq k|\mathcal{X},  p_k=t\}  &= \prod_{\substack{j\in\mathcal{X} \\ j\neq k}} \Pr\{p_j\leq t| \mathcal{X}\} \\
	&= \left[\frac{t}{p_{\pi(1)}}\right]^{|\mathcal{X}|-1}. 
\end{align*}
\begin{align}
&F_{S_{\pi(2)}}(x|\pi(1), p_{\pi(1)}, \mathcal{X}) \nonumber \\ &= \frac{1}{(p_{\pi(1)})^{|\mathcal{X}|}} \sum_{k\in\mathcal{X}} \int_0^{\min\left(F_{S_k}(x),~p_{\pi(1)}\right)} t^{|\mathcal{X}|-1} dt \nonumber \\
&= \frac{1}{|\mathcal{X}| (p_{\pi(1)})^{|\mathcal{X}|}} \sum_{k\in\mathcal{X}} \left[\min\left(F_{S_k}(x),~p_{\pi(1)}\right)\right]^{|\mathcal{X}|}. \label{eq:cond_SNR_2nduser_CUS}
\end{align}
Averaging (\ref{eq:cond_SNR_2nduser_CUS}) over $p_{\pi(1)}$ yields
\begin{align}
&F_{S_{\pi(2)}}(x|\pi(1), \mathcal{X}) \nonumber \\
	&= \frac{1}{|\mathcal{X}|} \int_0^1 u^{-|\mathcal{X}|} \sum_{k\in\mathcal{X}} \left[\min\left(F_{S_k}(x),~u\right)\right]^{|\mathcal{X}|} dF_{p_{\pi(1)}}(u) \nonumber \\
	&= \frac{1}{|\mathcal{X}|} \sum_{k\in\mathcal{X}} \left[ \int_0^{F_{S_k}(x)} \hspace{-0.8cm} dF_{p_{\pi(1)}}(u) + \hspace{-0.1cm} \int_{F_{S_k}(x)}^1 \hspace{-0.1cm}\left[\frac{F_{S_k}(x)}{u}\right]^{|\mathcal{X}|} \hspace{-0.3cm} dF{p_{\pi(1)}}(u) \right] \nonumber \\
	&= \frac{K}{|\mathcal{X}|} \sum_{k\in\mathcal{X}} \left[\int_0^{F_{S_k}(x)} \hspace{-0.6cm}  u^{K-1} du + \hspace{-0.1cm} \int_{F_{S_k}(x)}^1 \hspace{-0.1cm} [F_{S_k}(x)]^{|\mathcal{X}|} u^{K-1-|\mathcal{X}|} du \right] \nonumber \\
	&= \frac{K}{|\mathcal{X}|} \sum_{k\in\mathcal{X}} \left[ \frac{[F_{S_k}(x)]^K}{K} + \frac{[F_{S_k}(x)]^{|\mathcal{X}|} - [F_{S_k}(x)]^K}{K-|\mathcal{X}|} \right]. \label{eq:new_cond_SNR_2nduser_CUS}
\end{align}
We will next average (\ref{eq:new_cond_SNR_2nduser_CUS}) over $\mathcal{X}$. The probability of finding $m$ semi-orthogonal users to $\pi(1)$ under MU-Mode is given by 
\begin{align*}
	\Pr\{|\mathcal{X}| = m | \mbox{\small{MU-Mode}}\} = \frac{\alpha^{2m} (1-\alpha^2)^{K-1-m}}{1-\lambda}. 
\end{align*}
Let $\Omega_{m,k}(x) =  \left[\frac{[F_{S_k}(x)]^K}{K} + \frac{[F_{S_k}(x)]^m - [F_{S_k}(x)]^K}{K-m} \right]$. Averaging (\ref{eq:new_cond_SNR_2nduser_CUS}) over $\mathcal{X}$ gives 
\begin{align}
&F_{S_{\pi(2)}}(x| \pi(1)) \nonumber \\
&= \sum_{m=1}^{K-1} \frac{\alpha^{2m} (1-\alpha^2)^{K-1-m}}{1-\lambda} \frac{K}{m} \binom{K-2}{m-1} \sum_{\substack{k=1 \\ k\neq \pi(1)}}^{K} \Omega_{m,k}(x),\label{eq:pascal_triangle_explain} 
\end{align}
where (\ref{eq:pascal_triangle_explain}) comes from realizing that given $\pi(1)$ and $|\mathcal{X}|=m$, summing $\Omega_{m,k}(x)$ over all possible combinations of $m$ indices chosen from the set $\{1,\ldots,K\}\setminus \pi(1)$ is equivalent of summing over $\Omega_{m,k}(x)$ for all $k \neq \pi(1)$, $\binom{K-2}{m-1}$ times. 

Finally, we average (\ref{eq:pascal_triangle_explain}) over $\pi(1)$. Since each user is equally likely to be selected as $\pi(1)$, the distribution of $S_{\pi(2)}$ is finally given by
\begin{align} \label{eq:CDF_of_SNR_2nduser_CUS}
&F_{S_{\pi(2)}}(x) \nonumber \\
&= \sum_{i=1}^K \sum_{m=1}^{K-1} \frac{\alpha^{2m} (1-\alpha^2)^{K-1-m}}{1-\lambda} \binom{K-2}{m-1} \frac{1}{m} \sum_{\substack{k=1 \\ k\neq i}}^{K} \Omega_{m,k}(x)  \nonumber \\
								&= \sum_{m=1}^{K-1} c_m \sum_{k=1}^{K} \Omega_{m,k}(x).
\end{align}

By using (\ref{eq:CDF_of_SNR_2nduser_CUS}), we can calculate the ergodic rate of the second selected user in MU-Mode. The ergodic rate of the second selected user in MU-Mode using MRT can be written as 
\begin{align*}
\mathcal{R}_{M2}^{\mbox{\tiny{MRT}}}(\alpha) &{=} \sum_{m=1}^{K-1} c_m \sum_{k=1}^{K} \frac{\sigma^4}{g_k^2} \times \nonumber \\ & \hspace{0.2cm} \int_0^\infty \hspace{-0.2cm} \int_{0}^{\alpha^2} \hspace{-0.3cm} \log(1+(x^{-1}+y)^{-1}) dF_{Y^{\mbox{\tiny{MRT}}}}(y) dF_{S_{\pi(2)}}(x)  \nonumber \\
& {=} \sum_{m=1}^{K-1} c_m \sum_{k=1}^{K} \frac{\sigma^4}{g_k^2} \int_0^\infty \Upsilon^{\mbox{\tiny{MRT}}}(x) \Psi_{m,k}(x) dx, 
\end{align*}
and the ergodic rate of the second selected user in MU-Mode using ZF-BF can be calculated as
\begin{align*}
\mathcal{R}_{M2}^{\mbox{\tiny{ZF}}}(\alpha) &{=} \sum_{m=1}^{K-1} c_m \sum_{k=1}^{K} \frac{\sigma^4}{g_k^2} \times \nonumber \\ & \hspace{0.5cm} \int_0^\infty \int_{1-\alpha^2}^1 \log(1+xy) dF_{Y^{\mbox{\tiny{ZF}}}}(y) dF_{S_{\pi(2)}}(x) \nonumber \\
&{=} \sum_{m=1}^{K-1} c_m \sum_{k=1}^{K} \frac{\sigma^4}{g_k^2} \int_0^\infty \Upsilon^{\mbox{\tiny{ZF}}}(x) \Psi_{m,k}(x) dx,
\end{align*}
which completes the proof.

\balance

\bibliographystyle{IEEETran}
\bibliography{Tcom}

\end{document}